\documentclass[a4paper,twoside]{article}

\usepackage{epsfig}
\usepackage{subcaption}
\usepackage{calc}
\usepackage{amssymb}
\usepackage{amstext}
\usepackage{amsmath}
\usepackage{amsthm}
\usepackage{multicol}
\usepackage{pslatex}
\usepackage[bottom]{footmisc}

\usepackage{amssymb}
\usepackage{graphicx}
 \usepackage{balance}  
\usepackage{times}
\usepackage{algorithm}
\usepackage{algorithmic}
\usepackage{epsfig}
\usepackage{url}
\usepackage{latexsym}
\usepackage{color}
\usepackage{xspace}
\usepackage{arydshln}
\usepackage{amsmath}%
\usepackage{amsfonts}%
\usepackage{amssymb}%
\usepackage{setspace}
\usepackage{hhline}

\newtheorem{theorem}{Theorem}[section]
\newtheorem{lemma}[theorem]{Lemma}

\newtheorem{definition}{Definition}[section]

\usepackage{hyperref} 
\usepackage{apacite}
\usepackage{SCITEPRESS}     

\begin{document}

\newcommand{\SysName}{QTrail-DB }
\newcommand{\SysNameNS}{QTrail-DB}

\title{QTrail-DB: A  Query Processing Engine for Imperfect Databases with Evolving Qualities}

\author{\authorname{Maha Asiri, $~~$ Mohamed Y. Eltabakh}
\affiliation{Computer Science Department, Worcester Polytechnic Institute (WPI), MA, USA.}
\email{\{mmasiri,~meltabakh\}@wpi.edu}
}


\abstract{ Imperfect databases are very common in many applications due to various reasons 
ranging from data-entry errors,  transmission or integration errors, and wrong instruments' readings,
 to faulty experimental setups  leading to incorrect results.
The management and query processing of imperfect databases 
is a very challenging problem as it requires incorporating the data's qualities within the database engine.
Even more challenging, the qualities are typically not static and may evolve over time. 
Unfortunately, most of the state-of-art techniques deal with the data quality problem as 
an \textit{offline }task that is in \textit{total isolation} of the 
query processing engine  (carried out outside the DBMS). 
Hence, end-users will receive the queries' results 
with no clue on whether or not the results can be trusted for further analysis and decision making. In this paper, we propose the {it{``\SysNameNS''}} system that fundamentally extends the 
standard DBMSs to support imperfect databases with evolving qualities. 
\SysName introduces a new quality model based on the new concept of 
\textit{``Quality Trails''}, which captures the evolution of the data's qualities over time. 
\SysName extends the relational data model to incorporate the quality trails within the database system. 
We propose a new query algebra, called \textit{``QTrail Algebra''}, 
that enables seamless and transparent propagation and derivations 
of the data's qualities within a query pipeline. As a result, a query's answer 
will be automatically annotated with quality-related information at the tuple level. 
\SysName propagation model leverages the 
thoroughly-studied propagation semantics
present in the DB provenance and lineage tracking literature, 
and thus there is no need for developing a new query optimizer.
\SysName is developed within PostgreSQL and experimentally evaluated using 
real-world datasets to demonstrate its efficiency and practicality. }

\onecolumn \maketitle \normalsize \setcounter{footnote}{0} \vfill

\section{\uppercase{Introduction}}
\label{sec:introduction}

\label{sec:Intro}

In most modern applications it is almost a fact  that the working databases may  not  be perfect and may contain 
low-quality data records \cite{Batini2006, Redman:19025, sjkhdjk0879}.
The presence of such low-quality data is due to many reasons including missing or wrong values, 
redundant and conflicting information from multiple sources, human errors in data entry, 
machine and network transmission errors, or even wrong assumptions or instruments' calibration 
during scientific experimentations that lead to inaccurate results. 
A recent science survey has revealed that 80.3\% of the participant research and scientific groups 
have admitted that their working databases contain records of low quality, which puts their analysis 
and explorations at risk \cite{onlineScienceSurvy}. 
Moreover, it has been reported in \cite{Eckerson2000} that wrong decisions and uninformed analysis resulting from imperfect databases cost US businesses around 600 billion dollars each year.

\begin{figure}[t]
 \centering
   \includegraphics[height=7.7cm, width= 7.5cm, angle=0]{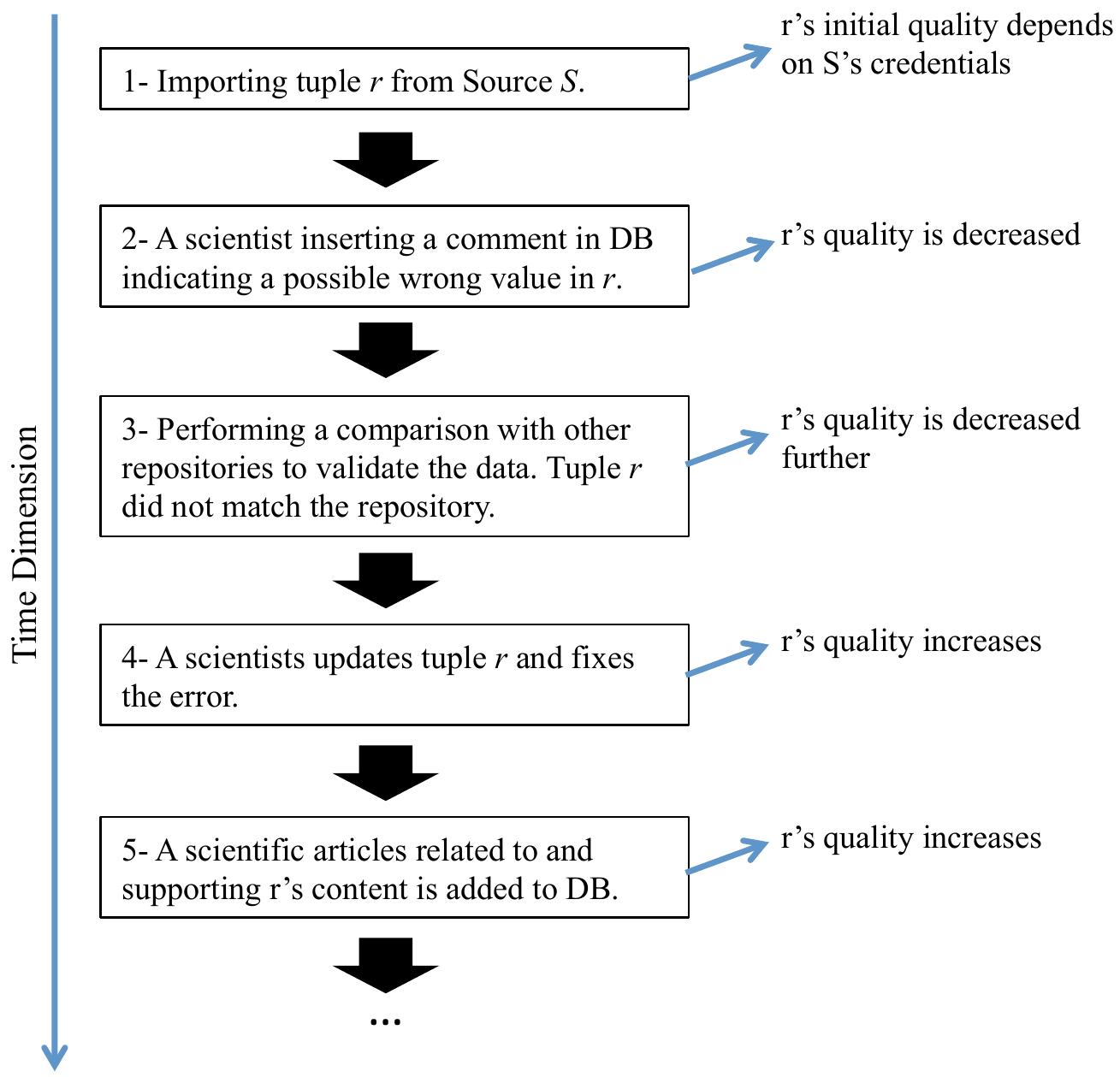}
   \caption{Database tuples with Evolving Qualities over Time.}
    \label{fig:motivation}
\end{figure}

Even more challenging, the qualities of the data tuples are typically not static, 
instead they may change over time ({\it{evolve}}) depending on various events taking place in the database. 
The emerging scientific applications are  excellent examples in which tracking and maintaining 
the data's qualities is of utmost importance~\cite{Radziwill2006, onlineScienceSurvy}.
 For example, Figure~\ref{fig:motivation}  illustrates  a possible sequence of operations that may take place 
in biological databases.  
First, a data tuple $r$ (e.g., a gene tuple) can be imported from an external source to the local database. 
At that time, $r$ would be assigned an initial quality score depending on the source's credibility.  
Then, a scientist may insert a comment highlighting a possible  error in the tuple (e.g., the gene's start position does not seem correct), 
based on which $r$'s quality should be decreased. 
After a while, a verification step that compares the local data with an external repository 
may confirm that $r$ contains an incorrect value, which will further decrease $r$'s quality. 
Subsequent actions in the database may either increase or decrease $r$'s quality over time, e.g., 
Steps 4 and 5 in Figure~\ref{fig:motivation}, which are an update operation on $r$ (e.g., correcting the gene's start position), and 
the addition of a  scientific article matching $r$'s new content, respectively, 
should both enhance $r$'s quality.   
In general, each tuple in the database may have its quality and trustworthy changing over time 
based on different operations taking place in the database.  

In such imperfect databases with dynamic and evolving qualities over time,
the standard query processing that treats all tuples the same while ignoring their qualities 
is indeed a very limited approach. For example, 
several interesting and challenging questions may arise beyond the standard data querying, which include:

\begin{enumerate}
    \item What was the quality of tuple $r$ before the last revision? 
    \item Why $r$'s quality has drastically dropped at time $t$, and what did we do to fix that?
    \item Given my complex query, e.g., involving selection, joins, grouping and aggregation, and set operators, what is quality of each output tuple? Can I trust the results and build further analysis on them or not? 
    \item Given my query, how to execute it  on only the high-quality tuples, e.g., the quality is above a certain threshold? How to join, select, or order the tuples based on their qualities?
    \item Among the low-quality tuples in the database, which ones are more important, e.g., frequently participate in queries' answers,  to investigate first?
\end{enumerate}

Certainly, supporting these types of  questions is of critical importance to end-users and high-level applications. 
On one hand, without modeling and keeping tracking of the quality information in a systematic way, crucial 
information will be lost.  
On the other hand, without assessing the quality of the output results, 
scientists and decision makers will become less confident about the obtained results.
And hence, building any further analysis on low-quality data may not 
only lead to wrong decisions, but also result in wasting scientists' efforts, resources, and budgets~\cite{Anders2011, Redman:19025}. 

It is clear that supporting these types of questions warrants the need for fundamental changes in the underlying DBMS. 
In this paper, we propose the {\it{``\SysNameNS''}} system, 
{\it{an advanced query processing engine for imperfect databases with evolving qualities}}.  
We identify three major tasks to be addressed, which are:

\subsection{Task 1$-$Systematic Modeling of Evolving Qualities}
With the large scale of modern databases, even a very small percentage of low-quality data may translate to a very large number of low-quality records. This makes it very challenging and time-consuming process to identify, isolate, or fix these records instantaneously. Therefore, the underlying database engine must be able to capture and model the data qualities in a systematic way, and also keep track of their evolution over time, e.g., when and why the quality changes (Refer to the aforementioned Questions 1 \& 2).  
\subsection{Task 2$-$Quality Propagation and Assessment of Query Results}		
It is impractical to assume that applications can freeze their working databases until all records have been fixed, and then enable them for querying. It is a continuous process of collecting and generating data of various degrees of qualities---with possible interleaving of offline efforts to verify and fix the imperfect tuples. Therefore, it is unavoidable to query the data while having  tuples of different qualities. Hence, the query processing engine must be extended to manipulate not only the data values but also their associated qualities. Each tuple $r$ in the output results should have an inferred and derived quality based on input tuples contributed to $r$'s computation (Refer to the aforementioned Question 3). 
	
\subsection{Task 3$-$Quality-Driven Processing and Curation}
Another important type of processing---beyond only propagating and deriving the output's quality---is the ability to query the data based on their qualities, i.e., {\it{quality-driven processing}}. 
This includes the ability to, for example, select, join, or order the data tuples based on their qualities, and possibly combine such quality-driven processing with the standard query operators in a single query plan (Refer to the aforementioned Question 4). Another type of analytics is the {\it{quality-driven curation}} in which end-users may want to, for example, track how low-quality tuples affect queries' results, or  rank the low-quality tuples according to their participation in queries for investigation and fixing purposes (Refer to the aforementioned Question 5).	Enabling this type of quality-driven processing mandates core changes in the database engine.

In this paper, we will focus on the first two tasks (Tasks 1 \& 2), while deferring Task 3 to future work. 
\SysName proposes a full integration of the data's qualities into all layers of a DBMS. This integration includes introducing a new quality model that captures the evolving qualities of each data tuple over time, called a {\it{``Quality Trail''}}, 
extending the relational data model to encompass the quality trails, and proposing a new relational algebra, called {\it{``QTrail Algebra''}}, that enables seamless and transparent propagation and derivations of the data's qualities within a query pipeline. 
Hence, each output tuple from a complex query plan will be annotated with its derived and inferred quality based on the contributing input tuples. Moreover, we show that \SysName extensions do not alter the query optimization rules, and hence the standard optimizers can be still used.  


The key contributions of this paper are summarized as follows:

\begin{itemize}
    \item Proposing the {\it{``\SysNameNS''}} system that treats data's qualities as
	an integral component within relational databases. In contrast to existing related work, e.g., the  offline quality management, and the generic provenance tracking systems, \SysName is the first to quantify and model the data's qualities, and fully integrate them within the data processing cycle. (Section~\ref{sec:related})
 
    \item Introducing a new quality model based on the new concept of {\it{``quality trails''}} that captures the evolving quality history of each data tuple over time. The new model enables advanced processing and querying over the quality trails, which is not possible in the  traditional single-score quality models. (Section~\ref{sec:model})
    
    \item Proposing a new query algebra, called {\it{``QTrail Algebra''}} that extends the semantics of the standard query operators to manipulate and propagate the quality trails in a pipelined and transparent fashion. Each operator annotates each of its output tuples with a quality trail that is inferred and derived from the contributing input tuples. We also propose new quality-specific operators and integrate them within the other operators. (Section~\ref{sec:propagation})

    \item Studying the viability of leveraging the powerful query optimizers of relational databases into \SysNameNS. We prove that under the new QTrail Algebra and by following the propagation semantics of provenance-based techniques, the standard equivalence and transformation rules are still applicable. (Section~\ref{sec:opt})

    \item Developing the \SysName prototype system within the PostgreSQL engine, and evaluating its performance using real-world biological datasets. Given the new value-added features, the results show acceptable overheads of \SysName compared to the standard query execution of traditional databases. (Sections~\ref{sec:creation} and~\ref{sec:exp})
 \end{itemize}

%
%


\section{\uppercase{Related Work}}
\label{sec:related}

Due to its critical importance, data quality has been extensively studied in literature. 
The most related to our work are the following.

\subsection{Cleaning and Repairing Technique}

A main thread of research is on data cleaning, repairing, and cleansing, where potential low-quality data records are identified, and then fixed~\cite{Bohannon175, DBLP:conf/icde/BohannonFGJK07, Cong:205890, Ebaid:2013:NGD:2536274.2536280, DBLP:conf/icde/LopatenkoB07,  Yakout:2012378}. The underlying techniques in these system vary significantly from fully-automated heuristics-based techniques~\cite{Bohannon175, Cong:205890, DBLP:conf/icde/LopatenkoB07}, comparison-based with external sources and repositories~\cite{DBLP:conf/icde/ArasuRS09, Beskale87695, Han:21663825}, and rule-driven techniques~\cite{DBLP:conf/icde/BohannonFGJK07, Bravo:2007:EDC:1325851.1325882}, to human-in-the-loop mechanisms~\cite{Galhardas033630, Yakout:2012378}. With the variety of algorithms and techniques for data cleaning, several extensible and generic frameworks have been proposed to integrate these algorithms, e.g.,~\cite{DBLP:confIOT13, Ebaid:2013:NGD:2536274.2536280, Geerts:2013:LDF:2536360.2536363}. The common theme in all of these systems is that they all work offline and in total isolation  from query processing. And since the data is evolving and growing rapidly in all modern applications, the repairing task is a never-ending time-consuming process. 

Therefore, it is inevitable that the data will be subject to querying, analysis, and decision making, while it contains low-quality records. Unfortunately, during query processing, none of the above techniques can provide any support for assessing the 
quality of the results or enabling quality-aware processing. Even worse, if these techniques have identified potential erroneous records and marked them as pending verification or fixing---which may take long time to complete, there is no mechanism to integrate such observations into query processing.

\subsection{Quality Assessment Techniques}

On the other hand, very little attention is given to quality assessment at query time. It has beed addressed in the context of mining operations~\cite{AyingModeling}, sensor data~\cite{Klein:2577845, Raymond2005}, and relational databases~\cite{DBLP:journals/tkde/BallouCW06, Motroingthe}. The core of  these techniques is based on statistical assumptions about the underlying data, e.g., defining statistical measures such as  {\it{completeness}}, {\it{soundness}}, and {\it{probability of error}}. And then, each technique studies its domain-specific operations and how they affect the statistical measures.
%

A major limitation in these systems is that the assumed statistics may not be available in many applications. For example, the work in~\cite{DBLP:journals/tkde/BallouCW06, Motroingthe}---which is the most related to \SysNameNS---assume that the probability of error in each column in the database is known in advance, which is not the case in many applications. And even if this knowledge is available, it a coarse-grained knowledge over an entire column and not tied to specific tuples, e.g., 1\% error rate in column $X$ means that among every 100 values in $X$, it is expected to find 1 error. Consequently, the estimated output quality is also coarse-grained and cannot be linked to specific tuples. Moreover, these systems assume a {\it{single-score}} quality model without taking into account the fact that data records are long-lived and their qualities evolve over time.  

\SysName is fundamentally different from the above mentioned two categories in that: (1)~It proposes a more rich quality model based on the quality trails instead of the single-score quality model, (2)~It does not put any assumptions on the data's characteristics, e.g., estimated error rate, instead the quality trails will be incrementally created and maintained as the database evolves over time, and (3)~Its quality model is fully integrated within the query processing engine.  

\SysName is complementary to the  cleaning and repairing techniques in that they together can provide
a more comprehensive solution that combines the online quality-aware query processing, and the offline repairing process, respectively. 
%

\subsection{Uncertain and Probabilistic Databases} 

Another big area of research is focusing on uncertain and probabilistic databases~\cite{p117,p115, ref78}. In these systems, a given data value (or an entire tuple) can be uncertain, and hence it is represented by a possible set of values, a probability distribution function over a given range, or a probability of actual presence. In  uncertain databases, the query engine is extended to operate on these uncertain values and tuples, and enforce correct semantics (called {\it{``possible worlds''}}).
Although uncertainty is related to data qualities in some sense, these systems are fundamentally different from \SysName since the notion of ``quality'' is not part of these systems. Therefore, the uncertain and probabilistic DBs can neither model or keep track of the data's qualities, nor enable advanced quality-driven query processing as proposed by the \SysName system.	

\subsection{Data Lineage and Provenance} 

Data provenance is directly related to data quality since the tuples' qualities are certainly based on their provenance. Several systems have addressed the derivation and propagation of the provenance information, e.g.,~\cite{ann-ref63, ann-ref66}, and even some systems such as Trio~\cite{ann-ref78} have combined the uncertainty with the provenance. 

\SysName  leverages the  thoroughly-studied  theoretical foundations and propagation semantics present in the provenance literature. However, there is still a big gap between capturing the raw provenance information, and the stage in which this information translates to quality-based knowledge and analytics.  For example, 

\begin{enumerate}
    \item Provenance systems can report the roots of each output tuple $r$---which may include 100s of other tuples' Ids---but they cannot inform scientists whether or not the results are of high quality and worthy of being used in further analysis. Currently, obtaining this knowledge involves a tremendous unrealistic manual effort  if at all possible.
    
    \item Provenance systems do not provide quantifiable measures on which queries can interact, i.e., lineage information are usually opaque objects with no easy way to apply conditions or transformations on. And
    
    \item Provenance systems do not capture the history (or evolution) of the data tuples, and thus executing the same query $Q$ at times $t_1$ and $t_2$ would return the same provenance information even if some changes between $t_1$ and $t_2$ have altered the records' qualities without changing their content, e.g., curation information is added that increase the virtue of the records. 
\end{enumerate}

\SysName addresses these issues and fills in this gap.


	 
\section{\uppercase{QTrail-DB DATA \& QUALITY MODELS}}
\label{sec:model}

\SysName has an extended data model, where each data tuple carries---in addition to its data values---a {\it{``quality trail''}} 
encoding the evolving quality of this tuple. More formally, for a given relation $R$ having $n$ data attributes, 
each data tuple $r \in R$ has the schema of: $r = \langle v_1, v_2, ..., v_n,  \mathcal{Q}_r\rangle$, where $v_1, v_2, ..., v_n$ are the data values of $r$, and $\mathcal{Q}_r$ is $r$'s quality trail.  $\mathcal{Q}_r$ is a vector in the form of $\mathcal{Q}_r =  \langle q_1, q_2, ..., q_z \rangle$, where each point $q_i$ is a quality transition defined as follows.

%

\begin{definition}[Quality Transition]
\label{def:transition}
{ A quality transition represents a change in a tuple's quality and it 
	consists of  a 4-ary vector $\langle$score, timestamp, triggeringEvent, statistics$\rangle$,  where  {\bf ``score''} is a quality score ranging between 1 (the lowest quality) and MaxQuality (the highest quality), {\bf ``timestamp''} is the time at which the score becomes applicable, {\bf ``triggeringEvent''} is a text field describing the event that triggered this quality transition,	and {\bf ``statistics''} field contains various statistics that will be maintained and updated during query processing.Only {\bf ``score''}, and {\bf ``timestamp''} are mandatory fields, while  {\bf ``triggeringEvent''}, and {\bf ``statistics''} are optional fields.
}
\end{definition}

\begin{figure}[t]
 \centering
   \includegraphics[height=4.9cm, width= 7.5cm, angle=0]{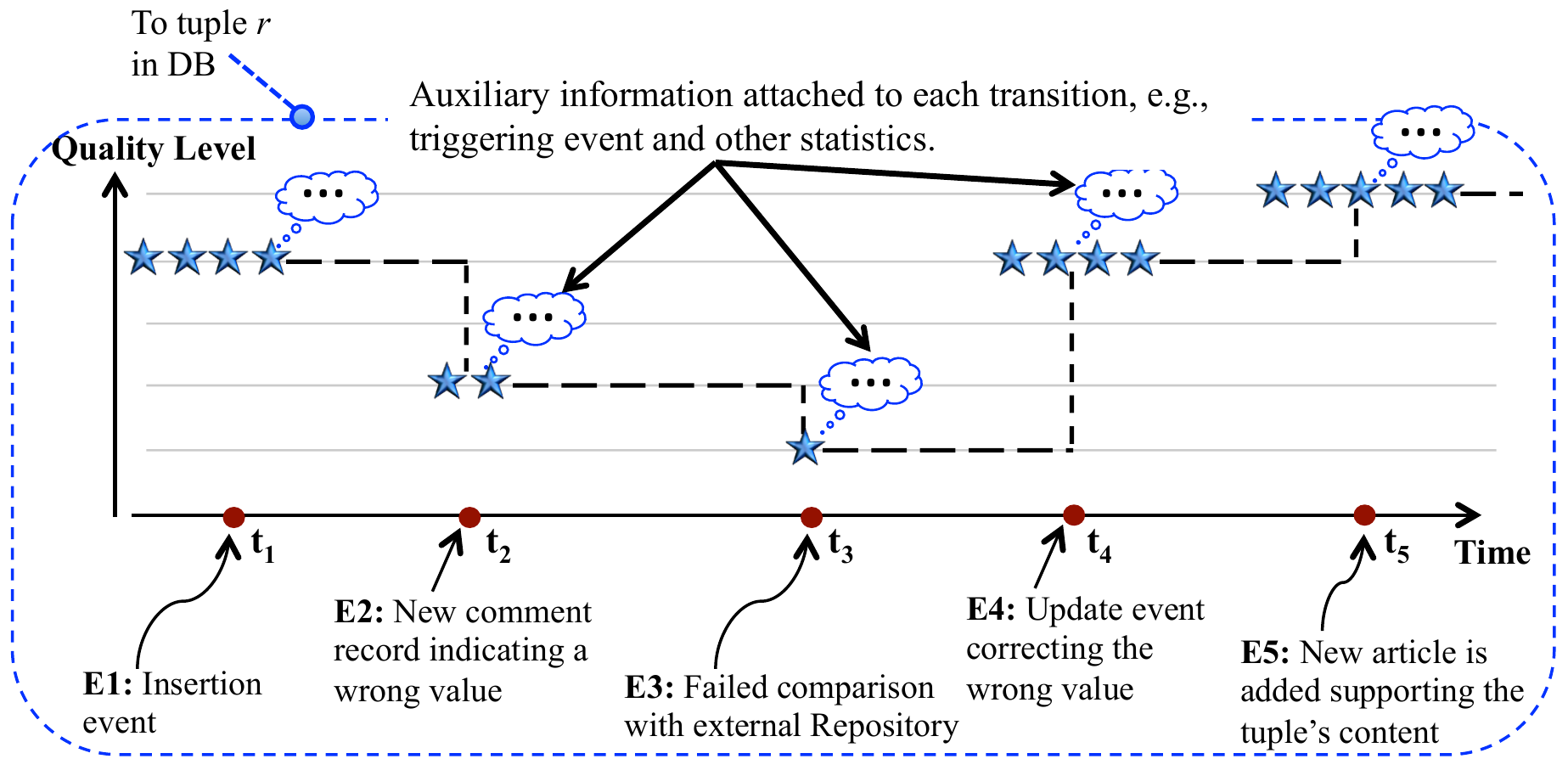}
   \caption{Example of $r$'s Quality Trail Corresponding To Operations in Figure~\ref{fig:motivation}.}
    \label{fig:model}
\end{figure}

Since $r$'s quality is evolving over time, the length of  $\mathcal{Q}_r$'s vector is also increasing over 
time by the addition of new transitions (Refer to  Figure~\ref{fig:model}). The quality trail is formally defined as follows.


\begin{definition}[Quality Trail]
\label{def:trail}
{ A quality trail of a given tuple $r \in R$ is denoted as $\mathcal{Q}_r$ and is represented as a vector of quality transitions. The transitions in $\mathcal{Q}_r$ are chronologically ordered, i.e., $\mathcal{Q}_r[i].timestamp < \mathcal{Q}_r[i+1].timestamp~\forall~i$. Moreover, the quality transitins have a stepwise changing pattern, 
 i.e., $\mathcal{Q}_r[i]$ is the valid transition over the time period [$\mathcal{Q}_r[i].timestamp, \mathcal{Q}_r[i+1].timestamp$). 
}
\end{definition}

Referring to the data tuple $r$ from Figure~\ref{fig:motivation}, its corresponding quality trail is depicted in  Figure~\ref{fig:model}. With each of the actions highlighted in Figure~\ref{fig:motivation}, $r$'s quality trail will change (evolve) from the L.H.S (the insertion time) to the R.H.S (the current time).  Each point in the quality trail is a quality transition. For example, at time $t_4$, a new quality transition is added to the trail consisting of: $\langle 4, t_4, ``updating~a~wrong~value", \{...\}\rangle$. This transition remains  valid  (the most recent one) until time $t_5$ when a new transition is added. The {\it{statistics}} field and its usage will be discussed in more detail in Section~\ref{sec:propagation}. 
  
\section{\uppercase{Quality Propagation and Assessment of Query Results}}
\label{sec:propagation}

In this section, we present the extended query processing engine of \SysName for propagating the quality trails within a query plan. To enable such derivation and propagation in a transparent and pipelined way, we propose a new SQL algebra, called {\it{``QTrail Algebra"}}, in which the standard query operators have been extended to seamlessly manipulate the quality trails associated with each tuple. In \SysNameNS, each operator will consume and produce tuples conforming to the data model presented in Section~\ref{sec:model}. In this section, we assume the quality trails have been created and maintained (The focus of Section~\ref{sec:creation}), and thus we will focus now on the query-time propagation.  

Fortunately, in the provenance literature, the propagation semantics of the tuples' lineage is a well studied problem under the 
different operators. In specific, we use the same semantics as in the Trio system~\cite{ann-ref78}. Therefore, after each algebraic transformation, we can track  the  input tuples contributing to a specific output tuple  without the need for re-inventing the wheel.  Yet, the unsolved  challenge is {\it{how to translate this knowledge to derivations over the quality trails?}} For example, assume that two tuples $r_1$ and $r_2$ will join together to produce an output tuple $r_o$, and from the existing literature, it is established that both $r_1$ and $r_2$ form the lineage of $r_o$. The remaining question is that given the quality trails of $r_1$ and $r_2$, what will be the derived quality trail of $r_o$? In the following, we study the semantics of deriving the quality trails of each output record from its contributing input records.

\subsection{Selection Operator ($\sigma_{p}(R)$)}

The operator applies data-based selection predicates $p$ over relation $R$, and reports the qualifying tuples. Predicates $p$ reference only the data values $v_1, v_2, ..., v_n$ within the tuples. The extension to the selection operator is straightforward since the content of the qualifying tuples do not change, and thus the output quality trails remain unchanged. The algebraic expression is:
$\sigma_p(R) = \{r = \langle v_1, v_2, ..., v_n, \mathcal{Q}_r\rangle \in R~|~ p(r) = True \}$ 

\subsection{Projection Operator $\pi_{a_1, a_2, ..., a_n}(R)$}

In \SysNameNS, the quality trails are at the tuple level, and not tied to specific attribute(s) within the tuple. Therefore, the projection operator will not change the quality of its input tuples. 
That is:
$\pi_{a_1, a_2, ..., a_n}(R) = \{r' = \langle a_1, a_2, ..., a_n, \mathcal{Q}_r\rangle\}$~$\forall~r \in R$.

\begin{figure}[t]
 \centering
   \includegraphics[height=6cm, width=7.5cm, angle=0]{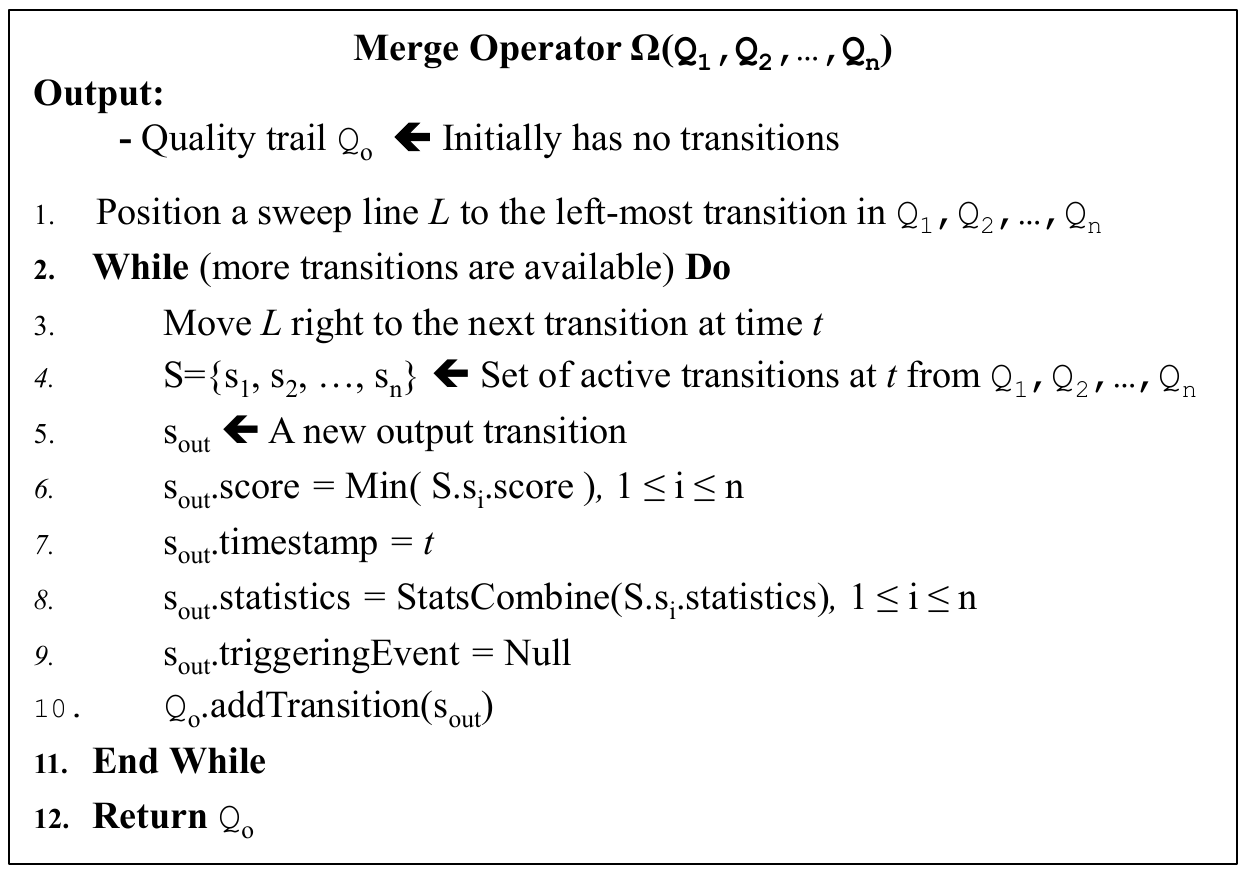}
   \caption{Pseudocode of the {\it{Merge}} $\Omega$ Operator.}
    \label{fig:MergeOp}
\end{figure}

\begin{figure*}[t]
 \centering
   \includegraphics[height=7.5cm, width= 15.5cm, angle=0]{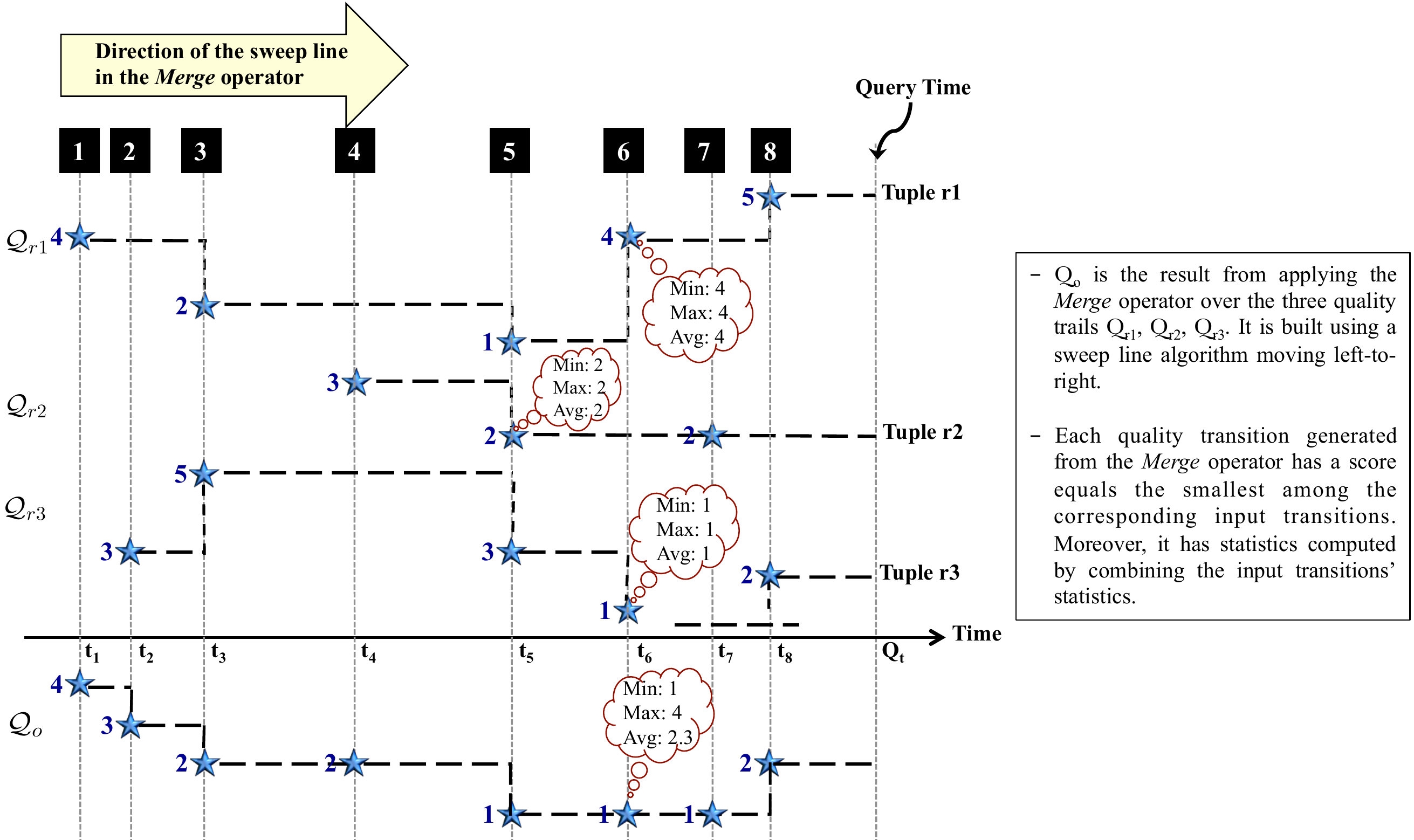}
   \caption{Example of the {\it{ Merge}} Operator in \SysNameNS.}
    \label{fig:op}
\end{figure*}

\subsection{Merge Operator ($\Omega(\mathcal{Q}_1, \mathcal{Q}_2, ...)$)}

Several of the relational operators, e.g., join, grouping, aggregation, among others, involve merging (or combining) multiple input tuples together to form one output tuple. Therefore, the corresponding input quality trails  may also need to be merged and combined together. To perform this merge operation over  quality trails, we introduce the new {\it{Merge}} operator $\Omega(\mathcal{Q}_1, \mathcal{Q}_2, ...)$. This operator  is not a physical operator, instead it is a logical operator that executes within other physical operators, e.g., join, grouping, and duplicate elimination.

The {\it{Merge}} operator's logic is presented in Figure~\ref{fig:MergeOp}, and its functionality is illustrated using the example in Figure~\ref{fig:op}. Assume combining three tuples $r_1$, $r_2$, and $r_3$ having quality trails  $\mathcal{Q}_{r1}, \mathcal{Q}_{r2},$ and $\mathcal{Q}_{r3}$, respectively. All quality trails are typically aligned from the R.H.S (which is the query time $Q_t$), i.e., each quality trial must have a valid transition at time $Q_t$. However, the trails are not necessarily aligned from the L.H.S since the data tuples may be inserted into the database at different times (See Figure~\ref{fig:op}). The quality trail of the output tuple  $\mathcal{Q}_{o}$ is derived using a {\it{sweep line}} algorithm over the input quality trails  starting from left to right and jumping over the transition points as illustrated in Figure~\ref{fig:op} (Lines 1-3 in Figure~\ref{fig:MergeOp}). The basic idea behind the algorithm is that the quality of the output tuple at any given point in time $t$  should be the lowest among the qualities of the contributing tuples at time $t$. This is based on the intuition that low-quality inputs produce low-quality outputs, and that an output tuple should have a high quality at time $t$ only if all its contributing input tuples have high qualities at $t$.  

Referring to the example in Figure~\ref{fig:op}, the sweep line starts at Position 1, where  only $\mathcal{Q}_{r1}$ exists and has a quality level {\it{4-star}}, which will be produced in the output. The line then jumps to Position 2, where $\mathcal{Q}_{r3}$ starts participating with a  quality level {\it{3-star}}, and hence a {\it{3-star}} transition will be added to $\mathcal{Q}_{o}$. The sweep line keeps moving to the subsequent positions,  and at each position, it calculates the lowest quality score among the input participants to be the output's quality score at this position (Lines 5-7 in Figure~\ref{fig:MergeOp}). For example, referring to the example in Figure~\ref{fig:op}(a), at time $t_4$, the contributing input qualities from $\mathcal{Q}_{r1}, \mathcal{Q}_{r2},$ and $\mathcal{Q}_{r3}$, are {\it{2-star}}, {\it{3-star}}, and {\it{5-star}}, and thus the corresponding quality transition on $\mathcal{Q}_{o}$ will have a {\it{2-star}} score. 

Although $\mathcal{Q}_{o}$'s quality scores reflect only  the lowest score among the input values,  the {\it{statistics}} field associated with each quality transition is intended to provide deeper insights on the other values contributing to the score. Initially, the statistics associated with each quality transition, e.g., {\tt Min}, {\tt Max}, {\tt Avg}, are set to the transition's score value as illustrated in Figure~\ref{fig:MergeOp}. And then, as the transitions get merged, new statistics are computed and get attached to the new quality transition. For example, the sweep line at Position 6 encounters scores {\it{4-star}}, {\it{2-star}}, and {\it{1-star}} transitions along with their initial statistics. Notice that $\mathcal{Q}_{r2}$'s active transition  at Position 6 is still the {\it{2-star}} transition occurred at Position 5 (at time $t_5$). These statistics will be combined by the {\it{Merge}} operator to compute the new statistics of the $\mathcal{Q}_{o}$'s new transition (Line 8 in Figure~\ref{fig:MergeOp}). The example illustrates maintaining the {\tt Min}, {\tt Max}, and {\tt Avg} statistics. However, in general, any type of aggregates that can be incrementally computed under the addition of new values, e.g., aggregates that are {\it{algebraic}}, e.g.,  {\tt Avg()} and {\tt Stddev()}, or {\it{distributive under insertion}}, e.g., {\tt Sum()}, {\tt Count()}, {\tt Min()}, {\tt Max()} can be supported~\cite{PalpanasSCP02}\footnote{Algebraic aggregates are those that need small or constant extra storage in order to be computed incrementally. Distributive-under-insertion aggregates are those that can be incrementally computed without extra storage when new data is added.}. 

Finally, the newly created transition is added to the output quality trail (Line 10 in Figure~\ref{fig:MergeOp}).

\subsection{Theta Join Operator ($R \Join_{p} S$)}

Given two tuples $r \in R$ and $s \in S$ having quality trails  $\mathcal{Q}_{r}$, and $\mathcal{Q}_{s}$, respectively. If $r$ and $s$ qualify for the data-based join predicate $p$ and produce a joint tuple $z$, then $z$ will inherit the merged qualities of its two components $r$ and $s$. The same intuition of the {\it{Merge}} operator applies where $z$'s quality at any given point in time $t$ should be the lowest among the contribution tuples. The algebraic expression is: 
\noindent
{\centerline{$R \Join_{p} S = \{\langle r.v_1, ..., r.v_n, s.v'_1, ..., s.v'_m,~\Omega(\mathcal{Q}_{r}, \mathcal{Q}_{s}) \rangle~|~$}}
\noindent
{\centerline{$ r \in R, s \in S, and~p(r, s) = True \}$}}

The Cartesian Product ($R~X~S$) and Natural Join ($R \Join S$) operators follow the same algebraic expression as the Theta Join operator. The only difference is that $p$ is {\it{True}} for the  Cartesian Product operator, whereas $p$ consists of equality predicates on the common attributes for the Natural Join operator.

The Outer Join operators (Left, Right, and Full outer joins) are direct extensions to the above-mentioned inner join operators. Without loss of generality, assume a tuple $r \in R$ will be in the output without a counterpart joining tuple from $S$, i.e., $S$'s values will be nulls, then $r$ will inherit its own quality trail. That is, $r$'s corresponding output tuple  will be: 

\noindent
{\small{$\{\langle r.v_1, ..., r.v_n, null, ...,null,~\mathcal{Q}_{r} \rangle\},~\forall~r \in R,~ \nexists~s \in S~|~p(r, s) = True$}}

\begin{figure*}[th]
 \centering
   \includegraphics[height=9cm, width= 15.5cm, angle=0]{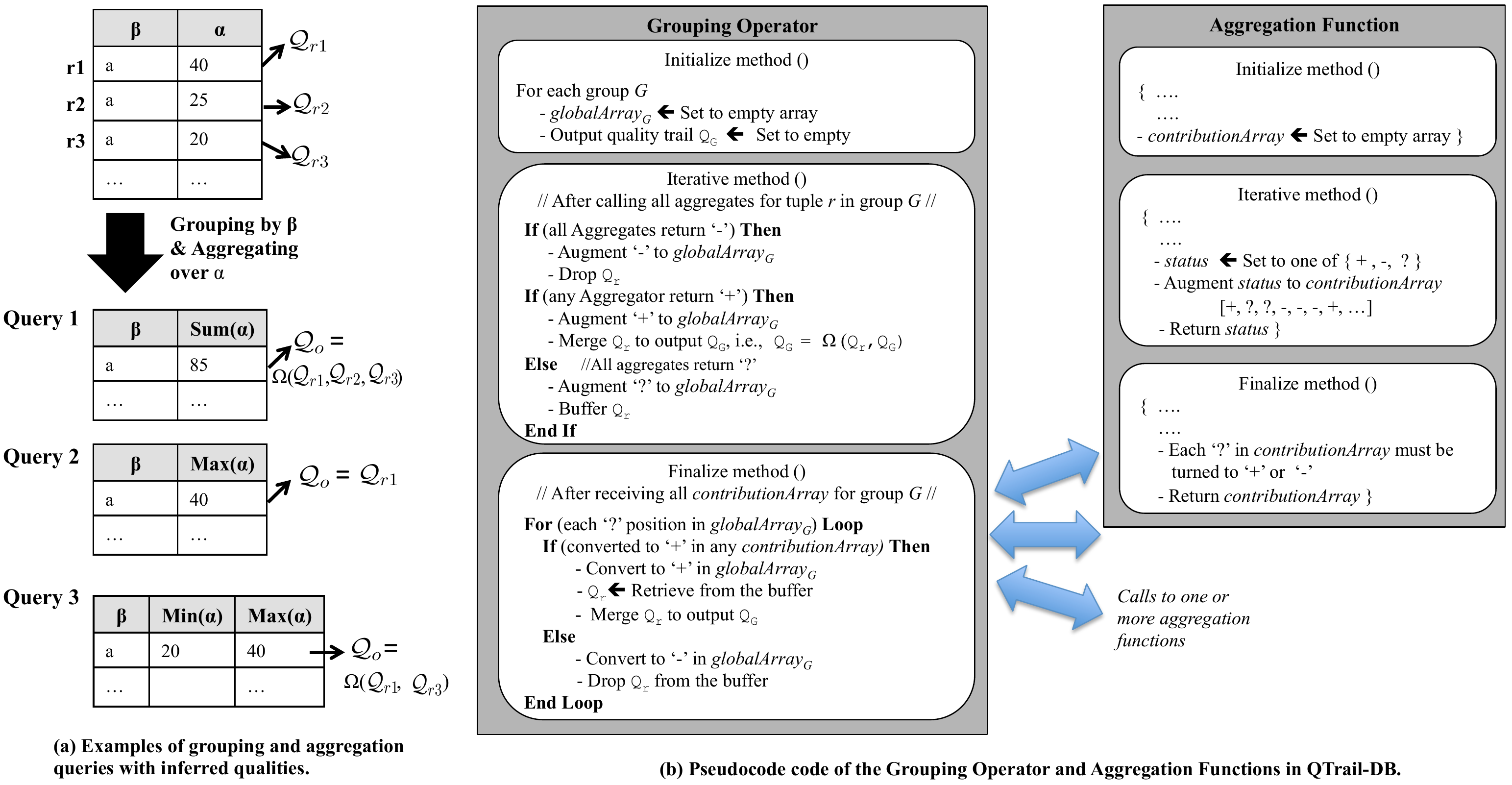}
   \caption{Extended Feedback Mechanism Between Grouping Operator and Aggregation Functions in \SysNameNS.}
    \label{fig:GroupingOp}
\end{figure*}

\subsection{Grouping and Aggregation ($\gamma_{a_1, a_2, ..., a_m, F_1(\beta_1), ...F_z(\beta_z)}(R)$)}

Grouping and aggregation are among the most interesting and challenging operators in \SysNameNS.	A key challenge lies in that in order to accurately infer the qualities of the output tuples, we need to keep track of the actual tuples participating in the aggregation. For example, for some aggregation functions only few tuples within each group  influence the output values. Referring to Figure~\ref{fig:GroupingOp}, assume that we group tuples $r_1$, $r_2$, and $r_3$ based on column $\beta$ and aggregate over column $\alpha$. The three illustrated queries in Figure~\ref{fig:GroupingOp}(a) compute different aggregation functions, i.e., {\tt Sum()}, {\tt Max()}, and \{ {\tt Min()} \& {\tt Max()}\}, respectively. Clearly, for the {\tt Sum()} function all of the tuples  $r_1$, $r_2$, and $r_3$ have contributed to the output tuple.In contrast, for the {\tt Max()} function, only $r_1$ contributed to the output, while in {\it{Query 3}} computing both the {\tt Min()} and {\tt Max()}, both $r_1$, and $r_3$ have contributed to the output.

In \SysNameNS, we support two semantics, which are the {\it{coarse-grained provenance-based semantics}}, and a new {\it{fine-grained}} semantics. In the coarse-grained semantics the aggregation functions are treated as black boxes, and hence all tuples within each group are assumed to participate in generating the output ({\it{black-box aggregation-unaware}} approach). In this case, all corresponding quality trails within each group will be merged together to derive the output's quality. This is the approach used in provenance-tracking Trio system~\cite{ann-ref78}. Although straightforward, this semantics is coarse grained and pessimistic, and may lead to under estimating the  outputs' qualities unnecessarily. Recall that, according to the {\it{Merge}} operator, as more quality trails are merged, the derived quality may only decrease and can never increase.  As these pessimistic derivations accumulate within a query plan, the final results may become useless and cannot be trusted because of its imprecisely-derived low quality.

\SysName implements supports also another {\it{open-box aggregation-aware}} semantics that identifies the minimal set of input tuples affecting the produced aggregation, and only those tuples are used to derive the output's quality. For example, as illustrated in Figure~\ref{fig:GroupingOp}(a), the different queries  will produce  different associated quality trails depending on the involved aggregation function, e.g., in Query 2, the output tuple has an assigned $\mathcal{Q}_{r1}$	 quality trail since only tuple $r1$ was involved. The merit of this semantics lies in avoiding unnecessary degradation of the output's qualities, and thus increasing the worthiness and usage of the produced results. However, implementing the open-box semantics requires extending the communication mechanism between the aggregation functions and the grouping operator as described next. 

In general, any aggregation function in \SysName---which relies on PostgreSQL DBMS~\cite{p106}---has three basic methods, i.e., {\it{Initialize()}}, {\it{Iterative()}}, and {\it{Finalize()}} (See Figure~\ref{fig:GroupingOp}(b)). The {\it{Initialize()}} function executes once before streaming the input records, {\it{Iterative()}}  is called for each input record to update the aggregator's internal state, and {\it{Finalize()}} executes once after all tuples have been processed to finalize the aggregator's output value. The proposed extension involves maintaining a character array, called {\it{contributionArray}}, by each aggregator function. And then, for each input tuple and based on the aggregator's semantics, the function decides on whether the tuple: (1)~Certainly contributes to the output, i.e., adding ``+'' to the array, (2)~Certainly does not contribute to the output, i.e., adding ``-'' to the array, or (3)~The tuple is in-doubt and it is not clear yet,  i.e., adding ``?'' to the array (See the R.H.S {\it{Iterative()}} method in Figure~\ref{fig:GroupingOp}(b)). In addition to accumulating each  tuple's status in the  {\it{contributionArray}}, this status is also returned to the grouping operator for an incremental update as we will discuss in sequel. Finally, the {\it{Finalize()}} method will return the  {\it{contributionArray}} to the grouping operator as illustrated in the figure.

While the aggregator function is  manipulating the {\it{contributionArray}}, \SysName enforces two invariants that the grouping operator will depend on to ensure correctness, which are (1)~The ``+'' and ``-'' values are permanent and cannot be altered, and (2)~By the time of returning the {\it{contributionArray}} from the aggregator's {\it{Finalize()}} function, the array  must not have any in-doubt tuples, i.e., any ``?'' entires must be converted to either ``+'' or ``-''. This is logical since the execution of the {\it{Finalize()}} method means that the aggregator has seen all its input tuples and knows exactly the ones actually contributed to the output. 

The grouping operator on the other hand---which may call multiple aggregation functions within a single query---maintains a global state across the different aggregation functions (See L.H.S of Figure~\ref{fig:GroupingOp}(b)). More specifically, the operator maintains a character array, called {\it{globalArray$_G$}} for each constructed  group $G$. And then, for a given input tuple $r$ belonging to $G$, if all aggregation functions return a ``-'' status, then $r$'s quality trail $\mathcal{Q}_{r}$ is immediately purged and ``-''  will be augmented to {\it{globalArray$_G$}}. This is because $r$ is certainly not contributing to $G$'s output. In contrast, if any aggregation function returns a ``+'' status, then $\mathcal{Q}_{r}$ is immediately merged into the quality trail of $G$'s output  $\mathcal{Q}_{G}$. Otherwise, $r$'s status is not decided yet, and hence a ``?'' will be augmented to {\it{globalArray$_G$}} and $\mathcal{Q}_{r}$ will be buffered by the grouping operator (See the Grouping operator's Iterative method in Figure~\ref{fig:GroupingOp}(b)). Finally, in the operator's {\it{Finalize()}} method, after all aggregation functions complete their work, the grouping operator will check the final status of the in-doubt tuples, say $r$. If any aggregation function alters its ``?'' status to ``+'', then the $r$'s quality trail will be retrieved from the buffer and merged into  $\mathcal{Q}_{G}$, otherwise $r$'s quality trail will be dropped.

The proposed extension of the grouping operator is internal to the database system and entirely transparent to the outside world. For the aggregation functions, we have extended the built-in functions, e.g., {\tt Min()}, {\tt Max()}, and {\tt Avg()}, to implement the new semantics. For new user-defined aggregation functions, the DB developer needs to implement the manipulation of the  {\it{contributionArray}} according to the function's semantics. In most aggregation functions that we have extended, 	this manipulation requires minimal effort, e.g., less than  10 additional lines of code. In return, the effect on the accuracy of the inferred qualities can be significant {\footnote{In the worst case, an aggregation function may blindly return ``+'' for each input, which will revert back to the black-box semantics.}}.

\subsection{Grouping Under Memory Constraints}

As presented in Figure~\ref{fig:GroupingOp}(b), the grouping operator may buffer the quality trail $\mathcal{Q}_r$ into memory if $r$'s status is not decided yet (The last two lines of the {\it{Iterative()}} method). However, for scalability, the grouping operator must be able to operate under limited memory, e.g., a memory buffer of max size $M$, without crashing. To achieve this and to avoid frequent writes to disk, which slows down the processing, we deploy an algorithm called {\it{``BufferClean()''}}.  This algorithm is executed  by the grouping operator when the allotted memory buffer is full. The functionality of BufferClean() is simple and involves retrieving the up-to-date {\it{contributionArray}} from the aggregation functions with the hope that the status of some of the buffered tuples has changed to either ``+'' or ``-''. In such case, their quality trails can be taken out from the buffer. In the worst case, if  {\it{BufferClean()}} did not free up some space, then the memory buffer is written to disk, and  will be later retrieved when the {\it{Finalize()}} method is called Interestingly, the  {\it{BufferClean()}}  algorithm can be very effective in many cases, and entirely avoids the need  to write to disk because several of the common aggregation functions that may require buffering, e.g., Min(), and Max(), may start discarding  previous candidates as they see few more tuples (See the experiments in Section~\ref{sec:exp}).

	
\subsection{Set Operators}	

The semantics of the set operators  $\cap$, $\cup$, and $-$ is extended in the same way as the other operators, i.e., if an output tuple is coming from both relations, then it will inherit the merged quality trails of the contributing input tuples. Otherwise, the output tuple is coming  from only one relation, and hence it will carry its own quality trail.The extended algebraic expressions of the set operators are defined as:

\vspace{1mm}
\noindent
$R \cap S = \{\langle z.v_1, ..., z.v_n,~\Omega(\mathcal{Q}_{R.z}, \mathcal{Q}_{S.z}) \rangle~|~ z~\in~R~\&~z~\in~S\}$
\noindent
$R - S = \{\langle r.v_1, ..., r.v_n,~\mathcal{Q}_{r}\rangle~|~ r~\in~R~\&~r~\notin~S\}$	
\[ R \cup S = \{ z~|~ z =\left\{ 
\begin{array}{ll}
\langle z.v_1, ..., z.v_n,~\Omega(\mathcal{Q}_{R.z},\mathcal{Q}_{S.z})\rangle&\text{if $C_1$}\\
 \langle z.v_1, ..., z.v_n,~\mathcal{Q}_{z}\rangle&\text{if $C_2$}
\end{array}\right\} \]

\noindent
where 

\vspace{1mm}
\noindent
$C_1 = (z~\in~R~\&~z~\in~S)$, and $C_2 = (z~\in~R~xor~z~\in~S)$

\subsection{Duplicate Elimination $\delta(R)$}

This operator involves eliminating duplicate tuples and keeping only one copy. In \SysNameNS, two tuples $r = \langle v_1, v_2, ..., v_n, \mathcal{Q}_r\rangle$ and $s = \langle z_1, z_2, ...., z_n, \mathcal{Q}_s\rangle$ are considered identical iff they are identical in the data part, i.e., $r.v_i = s.z_i,~  1 \leq i \leq n$. For each group $G$ of identical tuples, the quality trail of $G$'s output tuple is derived by merging the quality trails of $G$'s input tuples.

	
%
%
%

\section{\uppercase{Query Optimization in \SysName}}
\label{sec:opt}

A key advantage of building upon the propagation semantics in provenance literature such as the Trio system~\cite{ann-ref78}, i.e., the input tuples contributing to the quality of an output tuple $o$ are those that are considered $o$'s lineage, is the retention of the equivalence and optimization rules in relational databases. However, such retention cannot be directly claimed because some of the \SysName operators apply a complex merge function over the quality trails (See the {\it{ Merge}} operator $\Omega$ in Section~\ref{sec:propagation}). 
%
%
	Therefore, it is important to study the properties of this {\it{Merge}} operator,
	based on which  we can prove other algebraic properties.
	

\vspace{3mm}\noindent
\begin{lemma}[Properties of Merge Operator]
\label{lemma:merge}
The Merge operator has the following two properties:

~~~~{\it{{Commutativity:}}}~~~~~~~{\scriptsize{$\Omega(\mathcal{Q}_{1}, \mathcal{Q}_{2}$) = $\Omega(\mathcal{Q}_{2}, \mathcal{Q}_{1})$}}

~~~~{\it{{Associativity:}}}~~~~~~~{\scriptsize{$\Omega(\Omega(\mathcal{Q}_{1}, \mathcal{Q}_{2}), \mathcal{Q}_{3}) = 
	   								\Omega(\mathcal{Q}_{1}, \Omega(\mathcal{Q}_{2}, \mathcal{Q}_{3}))$ }} \\ 
{\scriptsize{$~~~~~~~~~~~~~~~~~~~~~~~~~~~~~~~~~~~~~~~~~~~~~~~~~~~~~~~~~~~~~~~~~~~~~~~~~~~~~~~=\Omega(\mathcal{Q}_{1}, \mathcal{Q}_{2}, \mathcal{Q}_{3}$)}}$\hfill\Box$

\vspace{3mm}\noindent
{\tt{Sketch of Proof:}}
	The proof of Commutativity is straightforward since
	the computation of each new transition (Lines 5-9 in Figure~\ref{fig:MergeOp}) 
	are not sensitive to the order of the input quality trails, e.g., $\mathcal{Q}_{1}$ and  $\mathcal{Q}_{2}$. 
	The Min() and StatsCombine() functions in Lines 6 and 8, respectively, treat their input values as a set, 
	and hence the order does not matter. 
	The proof of Associativity relies on that the primitive operations that generate each new transition from the input ones 
	are themselves associative. That is, the {\it{Min()}} function in Line 6 is associative, and also 
	{\it{StatsCombine()}} function, which combines the underlying statistics, involves associative functions. 
	Recall that the statistics need to be either {\it{algebraic}} (E.g., AVG and STDEV), or 
	 {\it{distributive under insertion}} (E.g., SUM, COUNT, MIN, MAX), and all of these functions are associative. 
\end{lemma}

Based on Lemma~\ref{lemma:merge}, we prove in the following theorem that 
the \SysNameNS's operators are order insensitive as it is the case in the standard relational operators.	
	
\vspace{3mm}\noindent
\begin{theorem}[Set Semantics]
\label{theorem:1}
Each operator in \SysName is guaranteed to generate the same quality-annotated output tuples independent of the inputs' order. $\hfill\Box$

\vspace{3mm}\noindent
{\tt{Sketch of Proof:}}	The standard relational operators are all order-insensitive as they execute under the Set or Bag semantics. And hence, they produce the same output independent of the inputs' order. The extended counterpart operators in \SysName can be categorized into three categories: (1)~Operators that do not apply any manipulation over the quality trails, e.g., $\sigma$, $\pi$, and $-$, (2)~Operators that apply the Merge operator to combine the quality trails, e.g., $\Join$, $\gamma$, $\cap$, and $\delta$, and (3)~Operators that fit in the two aforementioned cases, e.g., outer joins, and $\cup$. The operators in the $1^{st}$ category, by default, retain the order-insensitive property. The operators in the $2^{nd}$, and $3^{rd}$ categories are also guaranteed to infer the same quality for each output tuple independent of the inputs' order according to Lemma~\ref{lemma:merge}. Therefore, Theorem~\ref{theorem:1} is guaranteed to hold.
\end{theorem}

Theorem~\ref{theorem:1} is important as it guarantees the consistency of each operator's output independent of the inputs' order. 
One direct application of Theorem~\ref{theorem:1} is proving that the grouping operator 
will produce output tuples having the same inferred qualities 
despite the possible buffering of some quality trails until the 
end of the operator's execution (See the {\it{Iterative()}} function inside the grouping operator in Figure~\ref{fig:GroupingOp}).

\section{\uppercase{Creation \& Maintenance of Quality Trails}}
\label{sec:creation}

 \begin{figure}[t]
 \centering
   \includegraphics[height=6cm, width= 7.5cm, angle=0]{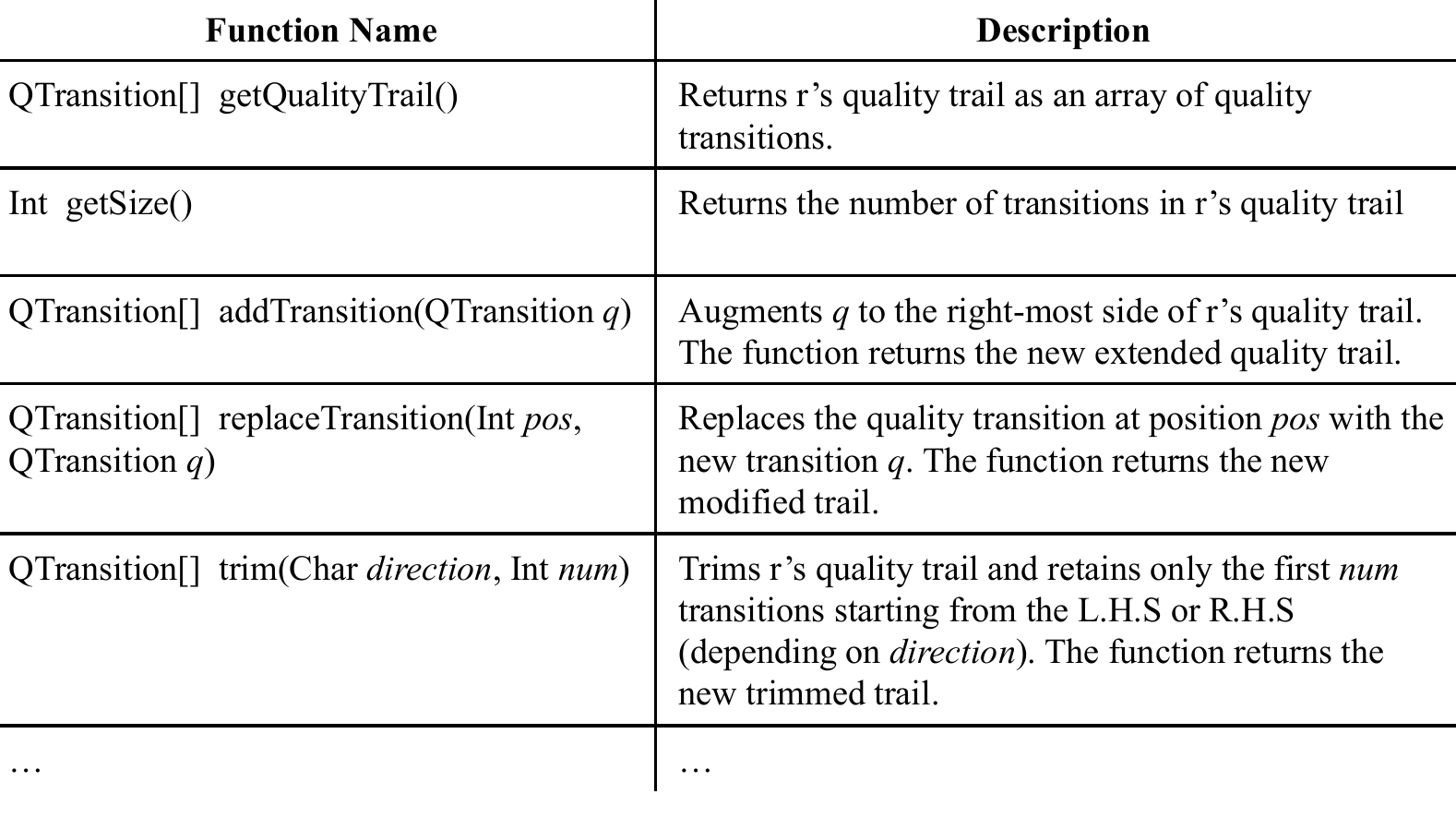}
   \caption{Manipulation Functions on $r.QTrail$ Attribute.}
   \label{fig:Func}
\end{figure}

In this section, we present the creation and maintenance mechanisms of the quality trails. 
Since \SysName is a generic engine, the goal is to design a set of APIs that will act as the interface between \SysName and 
the external world, e.g., the tools and applications' logics that can assess the quality scores and pass them to \SysNameNS. 
 More specifically, \SysName allows the database developers to manipulate the 
 quality trails as any other attribute in the database, e.g., modifying and extending quality trails  
in response to specific changes in the database from within the standard database triggers and UDFs. 
The quality trails are designed as special attributes added to the database relations, i.e., each 
 relation $R$ has an automatically-added special attribute called {\it{``QTrail''}}. 
 {\it{QTrail}} attribute is of a newly added user-defined type representing an 
 array of quality transitions (Refer to Definitions~\ref{def:transition}, and~\ref{def:trail}). 
 On top of this new type, a set of manipulation functions has been developed 
 as presented  in Figure~\ref{fig:Func}. These built-in functions are by no means comprehensive, 
 but they are basic functions on top of which 
the database developers may create more semantic-rich functions. 
 
 In Figure~\ref{fig:Func}, we present few of the developed functions.
  For example, assuming a given quality trail $r.QTrail$, the 
  {\tt getQualityTrail()} function returns the quality trail as an array of quality transitions, 
  while {\tt getSize()} function returns the number of transitions in the quality tail. 
  The functions {\tt addTransition()} and {\tt replaceTransition()} can be used to 
  alter an exiting quality trail by adding a new transition, or by modifying and replacing an existing transition, respectively. 
These functions internally double check that the timestamps of the transitions should be always monotonically increasing, 
otherwise the modification will be rejected.  
 The {\tt trim()} function enables trimming a given quality trail according to some 
 criteria, e.g., by specifying a direction (L.H.S or R.H.S), and a number of 
 transitions to keep. \SysName has several different signatures for the {\tt trim()} function. 
 This function is useful, especially if an application prefers not to keep the entire 
 history of the quality trail---which can be large, and 
 instead limit the scope to specific number of transitions or until specific time in the past. 
 In addition to these functions, we have also developed a set of function to manipulate 
 a given quality transition, e.g., building a quality transition, and setting  or retrieving specific fields within a transition{\footnote{Other storage schemes are possible without affecting 
the core functionalities of \SysNameNS. Only the implementation of the APIs may change.}}. 

 It is important to highlight that the tuples' quality trails are 
 exposed as normal attributes only for the maintenance and update purposes. 
 The objective from this feature is to make \SysName a generic engine with broader applicability to a wide range of application. 
However, this is entirely orthogonal to the quality propagation and assessment proposed 
 in Section~\ref{sec:propagation}. 
 First,  quality propagation does not require explicit querying 
 of the quality trails, i.e., users' queries will be written in the standard way as 
 in Example 1 in Section~\ref{sec:propagation}, and the quality trails will automatically propagate along with queries' results. 
 And second, the propagation semantics is too complex to be delegated to end-users and 
 to be encoded in each query.  That is why \SysName automatically and transparently 
 manage the propagation semantics.

\section{\uppercase{Experiments}}
\label{sec:exp}

{\textbf{Setup:}}
\SysName is implemented within the PostgreSQL DBMS~\cite{p106}. A Quality Trail is modeled as a 
new data type, i.e., a dynamic array of quality transitions. 
The default storage scheme is called {\it{``QTrail-Scheme''}}, in which 
each of the users' relations is automatically augmented with 
a new {\it{``QTrail''}} column as presented in Section~\ref{sec:creation}. 
\SysName is experimentally evaluated using an AMD Opteron Quadputer compute server with two 16-core
AMD CPUs, 128GB memory, and 2 TBs SATA hard drive. 
\SysName is compared with the plain PostgreSQL DBMS to 
study the overheads  associated with the new functionalities (w.r.t both time and storage).
No other work in literature offer functionaries similar to \SysName to compare with, 
especially that the work in~\cite{DBLP:journals/tkde/BallouCW06, Motroingthe} has not been experimentally evaluated.	

\begin{figure*}[t]
	\begin{minipage}[b]{0.61\linewidth}
	\centering
	\includegraphics[height=4.6cm, width= 9 cm, angle=0]{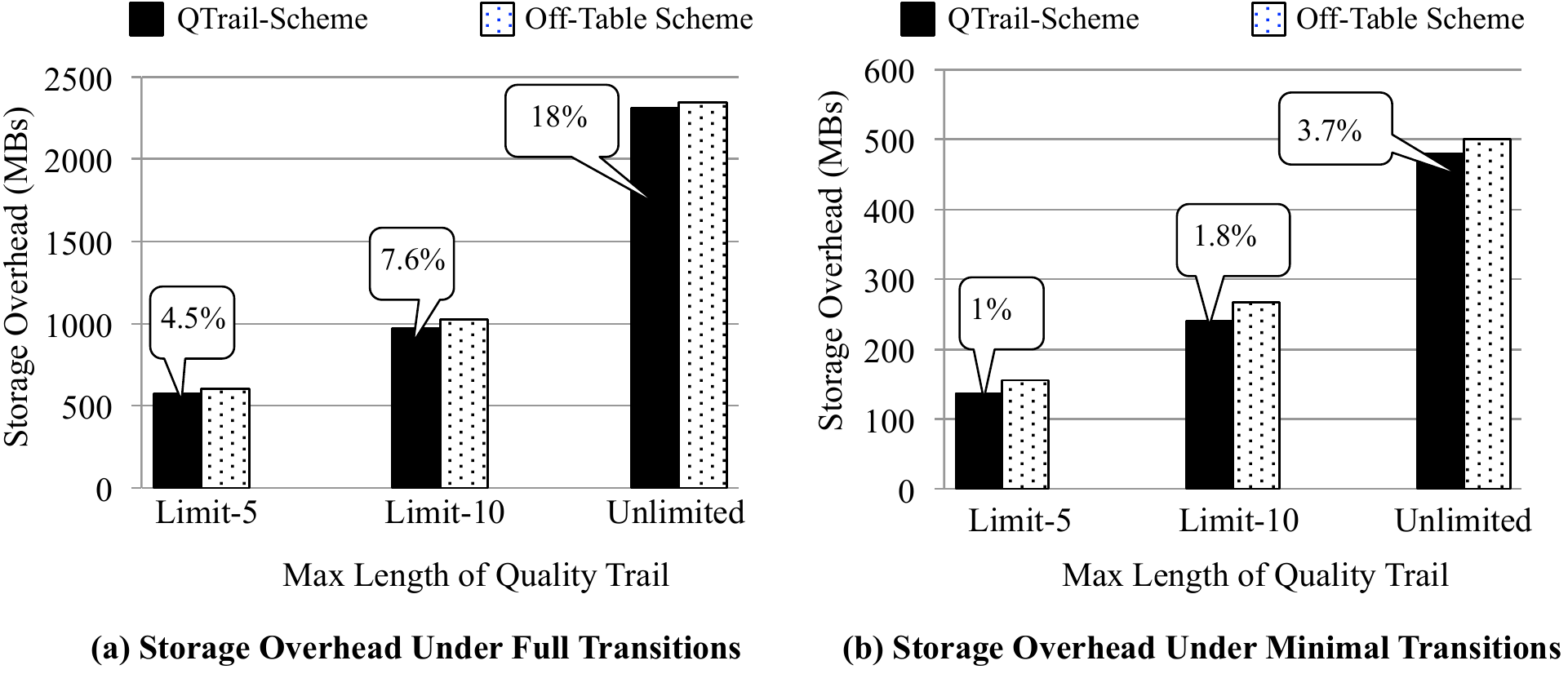}
	\caption{Quality Trail Storage Overhead.}
	\label{fig:storage}
	\end{minipage}
	\begin{minipage}[b]{0.39\linewidth}
	\centering
	\includegraphics[height=4.6cm, width= 6cm, angle=0]{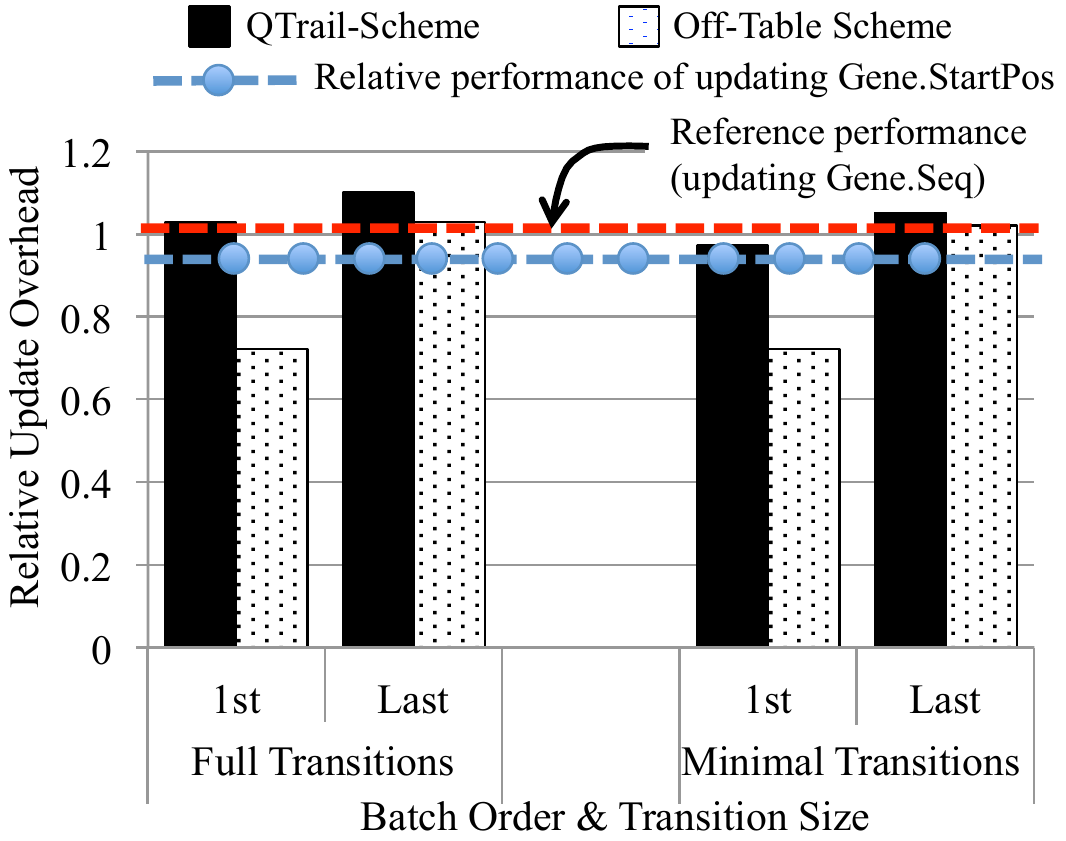}
	\caption{Quality Trail Update  Performance.}
	\label{fig:maintenance}
	\end{minipage}
\end{figure*}

\subsection{Application Datasets}

We use a subset of the curated UniProt real-world biological database~\cite{UniProt}.  UniProt offers a 
comprehensive repository for protein and functional information for various species. 
We extracted four main tables including {\tt Protein}, {\tt Gene}, {\tt Publication}, and {\tt Comment}.
The tables are connected through the following relationships:  The {\tt Protein} table 
has a many-to-one relationship with {\tt Gene}, many-to-many relationship with {\tt Publication}, and one-to-many relationship with {\tt Comment}. 
The {\tt Gene} table has also  a many-to-many relationship with {\it{Publication}}, and one-to-many relationship with {\tt Comment}.
The {\tt Comment} table contains free-text comments and notes related to the genes and proteins. 
These comments are extracted from the fields marked with ``CC" within the UniProt entires---which indicates a free-text comment field.  
Our dataset consists of approximately 750,000 protein records ($\approx$ 4.7GBs), 1.3 x 10$^6$ gene records ($\approx$ 8GBs), 
 12 x 10$^6$ publication records ($\approx$ 4.5GBs), and  8 x 10$^6$ comment records ($\approx$ 6.5GBs).
Thus, the total size of the dataset is approximately 24GBs.

\subsection{Workload}

We focus on tracking the qualities of the tuples in the {\tt Gene} and {\tt Protein} tables under the addition of new publications and comments. 
The quality score varies between $1$ (the lowest quality), and $10$ (the highest quality). 
To build the quality trails, we implemented an {\it{``After Insert"}} database trigger on each of the {\tt Publication} and {\tt Comment} tables
such that with the insertion of a new publication or comment, the quality of the corresponding gene or protein will be updated.  
We assume that the insertion of a new publication increases the 
quality (unless the quality score is already the maximum, in which case the new quality transition will have the same score as the previous one). 
For the comment values, each comment in UniProt has a code indicating the type of this comment. One 
of these types is {\it{``CAUTION''}}, which indicates a possible error or confusion in the data. 
All comments  having the {\it{``CAUTION''}} type are assumed to decrease the quality 
(unless the quality score is already the minimum, in which case the new quality transition will have the same score as the previous one). 
For the other comment values, we randomly labeled each one as ``+'', ``-'',  or ``$\sim$'', which 
indicates increasing, decreasing, or retaining the previous quality score, 
respectively~\footnote{We used random labeling since developing a free-text semantic extraction tool (or leveraging an existing tool)  is not the focus of this paper.}.  

Unless otherwise is specified, we assume the following: 
(1)~Each increase or decrease in the quality score is performed one step at a time, i.e.,  $\pm 1$, 
and (2)~If a quality transition is storing statistics---Referred to as {\it{``Full Transitions''}}--- then three types of statistics are maintained, 
	which are the \{{\tt Min}, {\tt Max}, {\tt Avg = (Sum, Count)}\}.
The insertions of the publication and comment records are randomly interleaved because the 
database does not maintain a global timestamp ordering the records' creation.
When evaluating the query performance of \SysNameNS, we compare against the standard query processing 
in which the quality trails are not even stored in the database.
Finally, the query optimizer of PostgreSQL has not been touched or modified, and hence the queries 
used in the evaluation  are optimized in the standard way.


\subsection{Storage and Maintenance Evaluation}

In Figure~\ref{fig:storage}, we study the storage overhead introduced by the quality trails. 
To put the comparison into perspective, we compare {\it{``QTrail-Scheme"}}  with another alternative 
where the quality trails of a given table $R$ are stored in a separate table $R${\it{-QTrail (OID, QTrail)}} 
that has one-to-one relationship with $R$. This scheme is referred to as {\it{``Off-Table Scheme"}}.
We study the storage overheads under the two 
cases of:~(1)~{\it{ Full Transitions}}, where each  quality transition has content in all its fields; the mandatory ones ({\it{score}}, and {\it{timestamp}}), 
and optional ones ({\it{triggeringEvent}}, and {\it{statistics}}) (Figure~\ref{fig:storage}(a)). 
In this case, the {\it{triggeringEvent}} field is a string of length varying between 50 and 100 bytes. 
And (2)~{\it{Minimal Transitions}}, where  each  quality transition has content in only  the mandatory fields (Figure~\ref{fig:storage}(b)). 

In each of the two figures, we measure the overhead under different  constraints on the maximum size of a quality trail (the x-axis). 
The values {\it{Limit-5}}, {\it{Limit-10}}, and {\it{Unlimited}} indicate keeping only up to the last 5, 10, or  $\infty$ transitions. 
The y-axis shows the absolute storage overhead, while the percentages inside the rectangle boxes show the overhead---more specifically that of  the QTrail-Scheme---as 
a percentage of the sum of {\tt Gene} and {\tt Protein}  sizes ($\approx$ 12.7GBs).
As Figures~\ref{fig:storage}(a) and~\ref{fig:storage}(b) show, there is no big difference between both storage schemes in all cases. 
The {\it{Off-Table}} scheme is slightly higher because of the storage of the unique tuple Id values ($OID$ column).  
In general, the quality trails do not introduce much storage overhead even under the worst case where the entire quality history is stored, e.g.,  
the overhead is around 18\% (for Full Transitions), and 3.7\% (for Minimal Transitions).
It is worth highlighting that under the {\it{Unlimited}} case, the longest quality trail consisted on 37 transitions.

In Figure~\ref{fig:maintenance}, we study the maintenance overhead of the quality trails. 
We consider the {\it{Unlimited}} case of quality trails. 
To have fair comparison, we measure the time of updating a quality trail, e.g., adding new transitions, 
w.r.t the time of updating other traditional fields, e.g., updating a text or integer fields. 
In all cases, the update operation takes place from within an {\it{After Insert}} trigger over the {\tt Comment} table as
described in the experimental workload above. Inside the trigger, the corresponding gene is retrieved  to update its data or its quality trail (See Example 2 in Section~\ref{sec:creation}). 
The retrieval from the DB uses a B-Tree index on the {\tt Gene.ID} column.  
Each operation is repeated 20 times, and their average is what we report. 
As illustrated in Figure~\ref{fig:maintenance}, we use the operation of updating a string field, more 
specifically replacing a segment of {\tt Gene.Seq} field with another segment, as our reference operation. That is, 
its execution time is normalized to value 1, and the other operations will be measured relative to this operation.  
The relative performance of updating an integer field, more specifically {\tt Gene.StartPos}, it also depicted in the figure. 
On average, the overhead of updating the integer field is around 94\% of updating the string field.  

For updating the quality trails, we consider both the {\it{QTrail-Scheme}}, and the {\it{Off-Table Scheme}}, 
and the two cases of {\it{Full Transition}}, and {\it{Minimal Transition}}. For each case, 
we run a batch that consists of 20 transactions inserting records into 
the {\tt Comment} table, which yield to updating the genes' quality trails.  
On the x-axis, we report the average performance over the entire batch under two scenarios: 
(1)~The batch is the $1^{st}$, i.e., all quality trails are empty, and (2)~The batch is the {\it{Last}}, where all other records have been 
inserted and the quality trails are almost complete. 
The results in Figure~\ref{fig:maintenance} show that operating on and updating a quality trail structure is very comparable to updating other database fields.
Under  the {\it{QTrail-Scheme}}, where the table to be queried and update is the {\tt Gene} table, 
the relative overhead ranged from 0.98\% (The $1^{st}$ batch with minimal transitions) to 1.11\% (The last batch with full transitions). 
The performance of the {\it{Off-Table Scheme}}, where the table to be queried and updated is called {\tt Gene-QTrail}, 
is almost the same except for the $1^{st}$ batch case, where the {\tt Gene-QTrail} table is empty.

 \begin{figure*}[t]
 \centering
   \includegraphics[height=4.5cm, width= 15.5cm, angle=0]{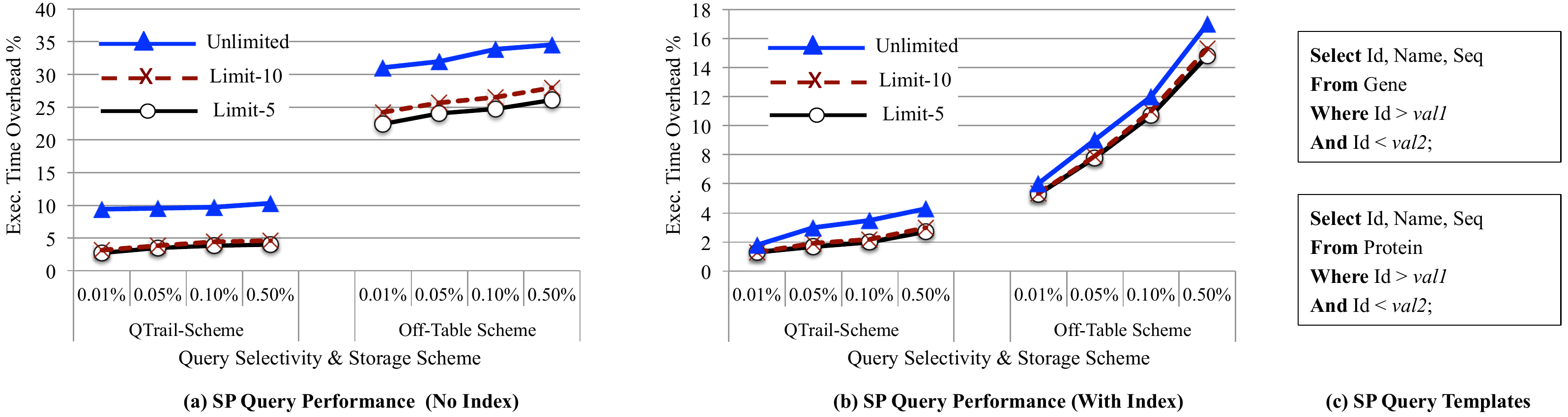}
   \caption{Performance of Select-Project (SP) Queries.}
    \label{fig:SP}
\end{figure*}

 \begin{figure*}[t]
 \centering
   \includegraphics[height=5cm, width= 15.5cm, angle=0]{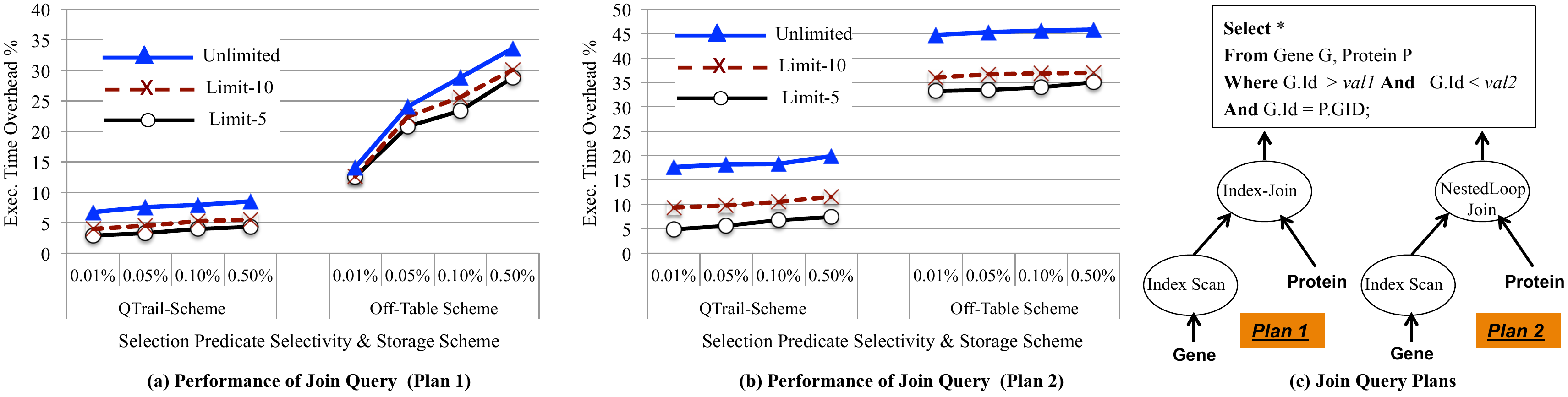}
   \caption{Performance of Join Queries.}
    \label{fig:join}
\end{figure*}


\subsection{SSP Query Performance}

In the next experiments, we evaluate the propagation and derivation of the quality trails at query time under various types of queries. 
In Figure~\ref{fig:SP}, we present the performance of Select-Project (SP) queries. The query templates are presented in  Figure~\ref{fig:SP}(c).
On the x-axis of Figures~\ref{fig:SP}(a) and~\ref{fig:SP}(b), we vary the query selectivity from 0.01\% (corresponds to 75 protein tuples, or 130 gene tuples) 
to 0.5\% (corresponds to 3,750 protein tuples, or 6,500 gene tuples),  and consider the two storage schemes {\it{QTrail-Scheme}} and {\it{Off-Table Scheme}}.
The y-axis measures the propagation overhead of the quality trails w.r.t the standard query 
performance, i.e., no quality trail storage or propagation.
We consider the propagation of the quality trails under the max size constraints of {\it{Limit-5}}, {\it{Limit-10}}, and {\it{Unlimited}}.
Under each configuration, we execute 5 queries on each of the {\tt Gene} and {\tt Protein} tables, and then report the average of the 
observed overheads across the 10 queries. 
Figures~\ref{fig:SP}(a) and~\ref{fig:SP}(b) show the results under the cases 
where the queries are executed without and with an index, respectively.  

The results show a big difference between the {\it{QTrail-Scheme}} and the {\it{Off-Table Scheme}}. 
This is mainly because the two operators that directly read from disk, i.e., the table-scan and index-scan operators, 
have different implementation under the two storage schemes. In the {\it{QTrail-Scheme}}, they are implemented such 
that they read the quality trails from the data tuples without any additional overhead. Whereas in the {\it{Off-Table Scheme}}, 
they are implemented to join the data tuples with the other table that contains the quality trails. 
All the other operators are independent of the physical storage of the quality trails as they read them from the operators' buffers in the query pipeline. 
Since join is an expensive operation, the  {\it{Off-Table Scheme}} encounters higher overhead. 
In Figure~\ref{fig:SP}(a), the query selectivity does not play a big factor because a complete table scan 
is performed regardless of the selectivity, which dominates most of the cost.
In contrast,  in Figure~\ref{fig:SP}(b), an index is used to select the data tuples satisfying 
the query's predicates, and thus the performance is more sensitive to the 
query selectivity. 
For the {\it{ff-Table Scheme}}, the overhead increases as the selectivity increases---and consequently the join cost increases. 
As the figure shows the {\it{Off-Table Scheme}} encounters between 2.5x and 6x higher overhead compared to the 
{\it{QTrail-Scheme}}.
It is important to highlight that the select and project operators do not apply any manipulation over the quality trails, and thus the encountered 
overheads are mostly due to the additional storage introduced by the quality trails.

\begin{figure*}[t]
 \centering
   \includegraphics[height=5cm, width= 15.5cm, angle=0]{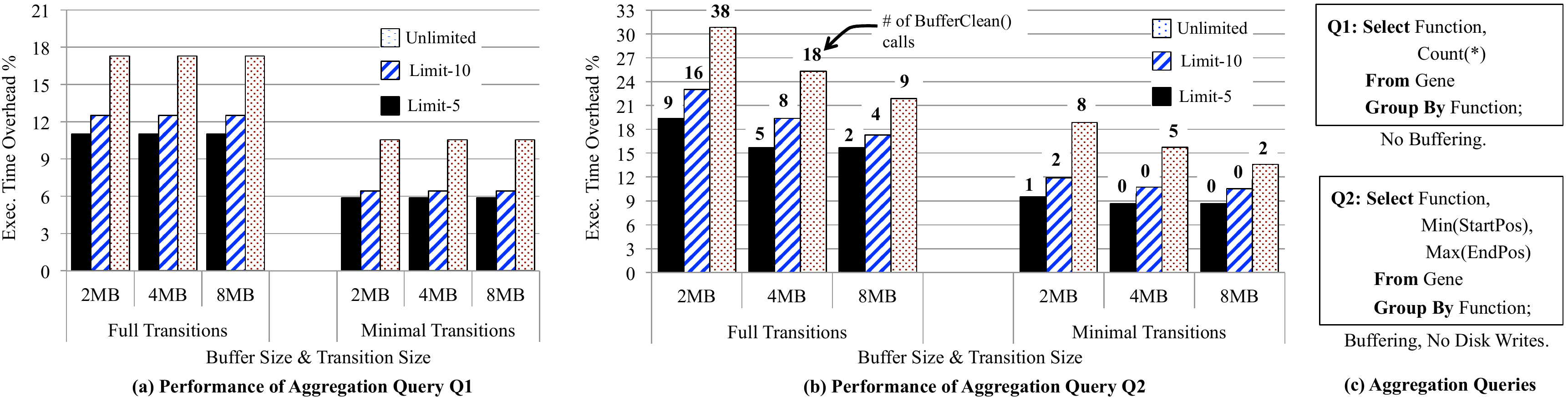}
   \caption{Performance of Aggregation Queries under the QTrail-Scheme.}
    \label{fig:agg1}
\end{figure*}

\subsection{Join Query Performance}

In Figure~\ref{fig:join}, we measure the performance of join queries. 
We consider joining the {\tt Gene} and {\tt Protein} tables using the query syntax highlighted in Figure~\ref{fig:join}(c).
Before the join operation, a selection predicated is applied over the {\tt Gene} table, and 
the selectivity of this predicates is varied between 0.01\% to 0.5\% as depicted on the x-axis of Figures~\ref{fig:join}(a), and~\ref{fig:join}(b).
The query is evaluated under two different query plans representing two different join types, i.e., 
in {\it{Plan 1}} the join operator uses an index on the the gene Ids  stored in the {\tt Protein} table ({\tt Protein.GID}), 
whereas in {\it{Plan 2}} such index does not exist, and thus the join is performed using block-based nested-loop algorithm. 
In Figure~\ref{fig:join}, we only report the results of the {\it{Full Transition}} case as  
the results of the {\it{ Minimal Transition}} case exhibit  similar trends, but with smaller overhead percentages.

In Figure~\ref{fig:join}(a), since both the selection and join operations are utilizing indexes, the quality trails' extra storage  
does not contribute much to the execution time overhead. 
That is why there is no big difference in performance under the different maximum sizes of 
quality trails ({\it{Limit-5}}, {\it{Limit-10}}, and {\it{Unlimited}}). 
In the case of the {\it{QTrail-Scheme}}, the execution 
overhead is mostly contributed to the merge operations applied over the quality trails within the join operator. 
Whereas, the {\it{Off-Table Scheme}} encounters the same overhead from the merge operations plus 
the additional join needed to link the data tuples to their quality trails, which is implicitly performed inside
the index-scan over both tables. This join overhead is clearly dominating the merge operations' overhead.
On the other hand, the performance results in Figure~\ref{fig:join}(b) are mostly not sensitive to the 
query's selectivity. This is because a full-table scan over the {\tt Protein} table will be performed in all cases, 
which dominates the overhead. That is also the reason why the {\it{Unlimited}} case shows 
a clear higher overhead compared to the {\it{Limit-5}} and {\it{Limit-10}} cases, i.e., the difference in I/Os
becomes a distinguishing factor among the three cases.

\subsection{Aggregation Query Performance}

For the grouping queries, the type of the aggregation function(s) plays an important role. 
This is because some aggregation functions, e.g., COUNT, SUM, AVG,  do not require any buffering for the input quality trails, 
while other functions, e.g., MIN, and MAX, will most probably require buffering. 
In Figure~\ref{fig:agg1}, we study the aggregation performance under these two types of aggregators (Figure~\ref{fig:agg1}(c)). 
Query $Q1$ involves one function COUNT(), which always returns ``+'' for each input tuple, and hence the quality trails are incrementally merged. 
In contrast, Query $Q2$ involves two functions MIN() and MAX(), and each may return ``-'' if the tuple is certainly not contributing to the function's result, 
or ``?'' if the tuple may contribute. Since the minimum and maximum values cannot be determined until the last tuple is seen, then the quality trails of the  
in-doubt tuples will need to be buffered. In our implementation of the MIN() (or MAX()) functions, if multiple tuples have the same 
minimum (or maximum) value, then they all contributing to the output result.
In this set of experiments, we will focus only on the {\it{QTrail-Scheme}} since it has proven superiority over the {\it{Off-Table Scheme}}.
 
The results in  Figure~\ref{fig:agg1}(a) show that changing the maximum buffer size (the x-axis) has no effect on the performance since  COUNT(*) does not require buffering.  
Nevertheless, the transition size (either {\it{Full}} or {\it{Minimal}}) and the maximum 
allowed size of a quality trail (either {\it{Limit-5}}, {\it{Limit-10}}, or {\it{Unlimited}}), are both affecting the performance. 
This is because the grouping query requires a full table scan, and thus as the storage overhead of the quality trails increases, the execution time also increases.  
The storage overhead also plays the same effect in Figure~\ref{fig:agg1}(b). However, $Q2$ does buffering, and thus the allowed buffer size affects the performance. 
As presented in Section~\ref{sec:propagation}, when the buffer is full, the grouping operator will execute the {\it{BufferClean()}} function trying to avoid un-necessary disk-writes, and 
get rid of the in-doubt tuples (``?'' ) that have changed their status to not-contributing (``-'').  
Figure~\ref{fig:agg1}(b) shows the number of  {\it{BufferClean()}} calls in each case. 
For example, for the minimal-size transitions with 8MBs buffer size, only two {\it{BufferClean()}} calls are 
executed for the {\it{Unlimited}} case, and zero calls for the {\it{Limit-5}} and {\it{Limit-10}} cases.
For this experiment, these  {\it{BufferClean()}} calls are very effective since the in-doubt tuples can be resolved fast as more tuples are seen, and thus no disk writes are 
required even under the smallest buffer size of 2MBs. Therefore, depending on the different configurations, the relative execution overhead (w.r.t the standard query) 
varies between 8\% and 31\% as illustrated in the figure.

\begin{figure}
 \centering
   \includegraphics[height=4.5cm, width= 7.5cm, angle=0]{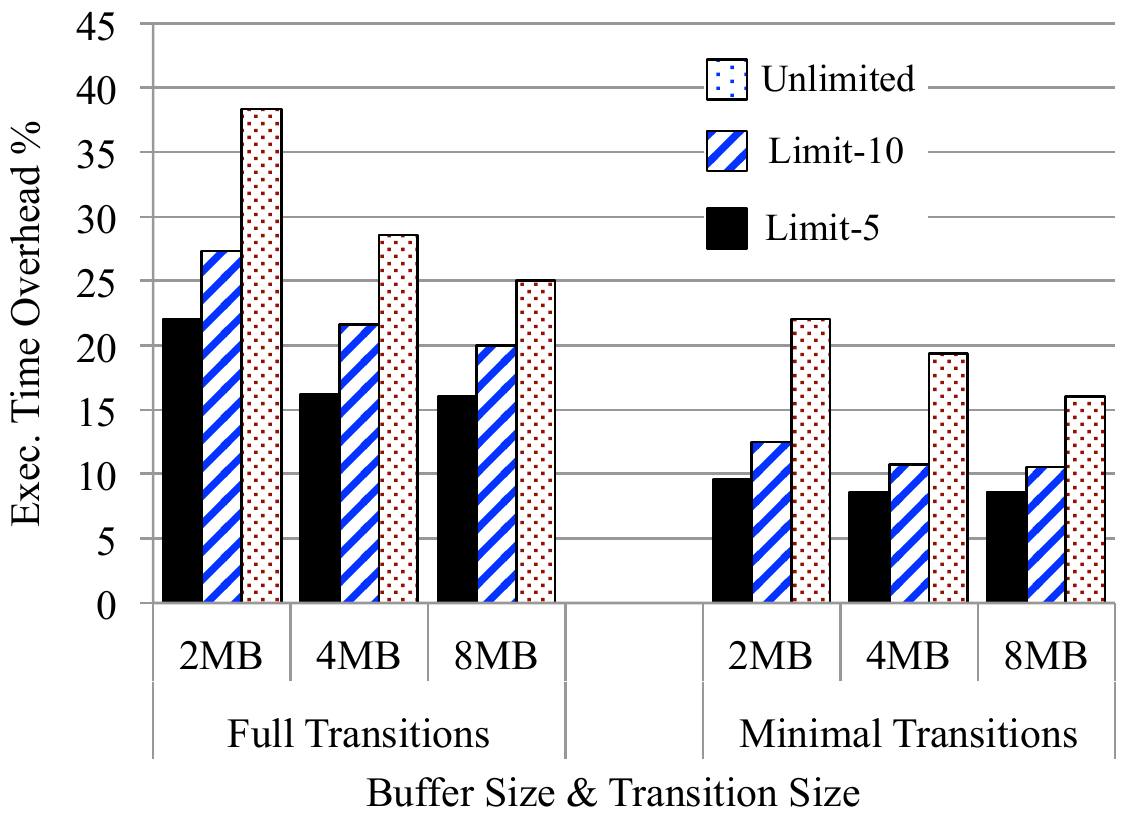}
   \caption{Aggregation Query Q2: Disabled BufferClean() and Enabled Disk-Writes.}
   \label{fig:agg3}
\end{figure}

In some aggregation function, the execution of {\it{BufferClean()}} may not be always effective, and thus the grouping operator will need to write the buffer's content to disk. 
To study such effect, in Figure~\ref{fig:agg3}, we repeat the same experiment of Query $Q2$ reported in Figure~\ref{fig:agg1}(b), but in this case we disable the 
{\it{BufferClean()}} function. And hence, the grouping operator has to write the buffer to disk when full. 
The results in Figure~\ref{fig:agg3} confirm the same trend as the other experiment 
with the exception of having a higher execution overhead (between 0\% in the best case to 6.5\% in the worst case). 
It is worth mentioning that if an aggregation function requires extensive buffering for the quality trails, e.g., the MEDIAN() function would buffer all its input quality trails, 
then most probably this function would also require extensive buffering of the data 
values. In this case and for performance considerations, these functions can easily implement the 
coarse-grained semantics, i.e., returning ``+'' for each input tuple, and hence no quality trail buffering is needed. 

\begin{figure}
 \centering
   \includegraphics[height=4.5cm, width= 7.5cm, angle=0]{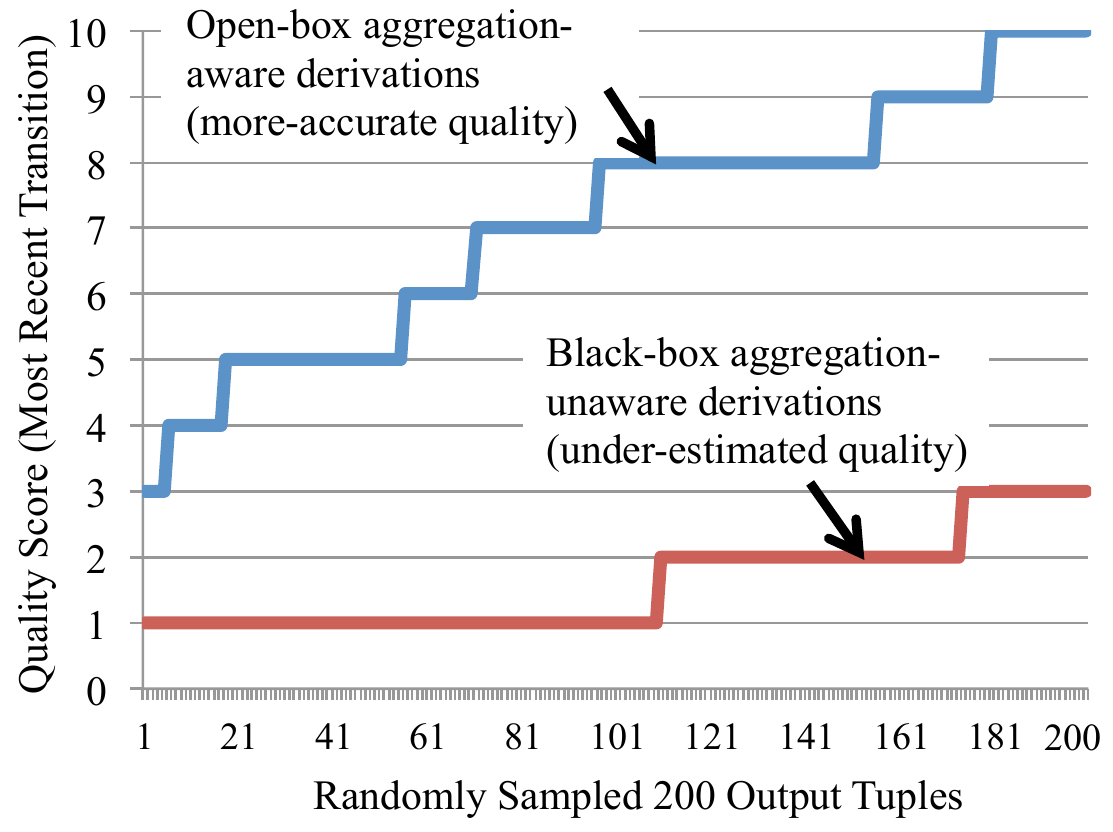}
   \caption{Aggregation Query Q2: Quality Score under Coarse- vs. Fine-Grained Derivation.}
    \label{fig:agg4}
\end{figure}

To illustrate the benefits from implementing the {\it{open-box aggregation-aware}} quality derivation  for the grouping and aggregation operators,
we repeat the aggregation query $Q2$ under the {\it{black-box aggregation-unaware}} implementation, i.e., all input tuples within a group 
contribute to the output's quality of the {\tt Min()} and {\tt Max()} functions (Figure~\ref{fig:agg4}). 
We randomly sampled 200 tuples from the query's output, and report in Figure~\ref{fig:agg4}, 
the quality score of the most recent transition for these tuples (in a sorted order). 
The  results confirm that the black-box approach can severely and unnecessarily under estimate  the tuples' qualities.  
For example, in the open-box approach, most tuples have quality scores higher than 6. 
Whereas, in the black-box approach, most tuples have the lowest quality score of 1.

\section{\uppercase{Conclusion}}
\label{sec:Conclusion}
We proposed  \SysName  as an advanced query processing engine for imperfect databases with evolving qualities. 
At the conceptual level,  \SysName enables  high-level applications to
model their data's qualities inside the database system, keep track of how the qualities evolve over time, 
and build more informed decisions based on the automatically quality-annotated query results.   
At the technical level, \SysName involves several novel  contributions including:
(1)~Introducing a new quality model based on the new concept of {\it{``quality trails''}} in contrast to the commonly-used {\it{single-score}} quality model, 
(2)~Extending the relational data model to include the quality trails, 
(3)~Proposing a new query algebra, called {\it{``QTrail Algebra''}}, which extends the standard query operators as well as 
	introduces new quality-related operators for the propagation and derivation of quality trails at query time, 
and (4)~Proving that \SysNameNS's query optimizer can inherit and leverage the logic of the standard query optimizers 
	without the need for customized optimizers. 
The experimental evaluation has shown the practicality of \SysNameNS, and the efficiency of its design choices.


{\small
\bibliography{Ref/quality,Ref/uncertain,Ref/ann}}

\end{document}